%% file: main.tex
\documentclass[sigplan,screen,10pt]{acmart}
\usepackage{soul, color, colortbl}
\usepackage{hyperref}
\usepackage{hyperxmp}

\usepackage{amsmath}
\usepackage{algorithm}
\usepackage{tabularx}
\usepackage{makecell}
\usepackage{multirow}
\usepackage{multicol}
\usepackage{tabularx}
\usepackage{xcolor}
\usepackage{hhline,booktabs,amsmath}
\usepackage[noend]{algpseudocode}
\definecolor{mygray}{gray}{0.9}
\definecolor{dkgreen}{rgb}{0,0.6,0}
\definecolor{mauve}{rgb}{0.58,0,0.82}
\definecolor{mygray}{gray}{0.9}
\newcommand{\sol}{{WaterWise}}
\newcommand{\todo}[1]{\textcolor{black}{#1}}
\setcopyright{none}
\renewcommand\footnotetextcopyrightpermission[1]{}
\author{\text{Yankai Jiang}}
\affiliation{
    \institution{Northeastern University}
    \country{}
}

\author{\text{Rohan Basu Roy}}
\affiliation{
    \institution{University of Utah}
    \country{}
}
\author{\text{Raghavendra Kanakagiri}}
\affiliation{
    \institution{IIT Tirupati}
    \country{}
}

\author{\text{Devesh Tiwari}}
\affiliation{
   \institution{Northeastern University}
   \country{}
}

\begin{document}

\title[\sol{}: Environmentally Sustainable Large-Scale Computing]{\sol{}: Co-optimizing Carbon- and Water-Footprint Toward Environmentally Sustainable Cloud Computing}

\begin{abstract}
\textit{The carbon and water footprint of large-scale computing systems poses serious environmental sustainability risks. In this study, we discover that, unfortunately, carbon and water sustainability are at odds with each other -- and, optimizing one alone hurts the other. Toward that goal, we introduce, \sol{}, a novel job scheduler for parallel workloads that intelligently co-optimizes carbon and water footprint to improve the sustainability of geographically distributed data centers.}
\vspace{-1mm}
\end{abstract}

\ccsdesc[500]{Social and professional topics~Sustainability}
\ccsdesc[500]{Computer systems organization~Cloud computing}

\keywords{Cloud Computing, Sustainability, Geospatial Shifting}

\maketitle

\input{sections/introduction}
\input{sections/background}
\input{sections/motivations}
\input{sections/design}
\input{sections/methodology}
\input{sections/evaluation}
\input{sections/discussion}

\input{sections/related_work}

\input{sections/conclusion}

\bibliographystyle{ACM-Reference-Format}
\bibliography{refs}
\clearpage
\input{sections/artifact}

\end{document}

%% file: sections/introduction.tex
\section{Introduction}
\label{sec:intro}

\noindent\textbf{Problem Motivation and Insights.} The rapidly increasing carbon footprint of large-scale computing systems is posing serious environmental sustainability risks -- and, has consequently received significant attention from the computer systems research community~\cite{gupta2021chasing,gupta2022act,eeckhout2022first,li2023toward,wu2022sustainable,acun2023carbon, 2024limitationssukprasert,wang2024designing, hanafy2024gaia,anderson2023treehouse}. Unfortunately, the environmental impact of computing systems on another critical environmental sustainability factor --- water footprint -- has not been well-explored. A recent study has shown the dramatic side effects that parallel computing workloads, including ML training, place on our water resources~\cite{li2023making}. Mitigating the water footprint challenge is critical to making large-scale cloud computing more sustainable. Without a focus on the water footprint of the computing infrastructure, we cannot achieve holistic sustainability of large-scale cloud computing systems.

\vspace{2mm}
Motivated by these factors, in this study, we investigate the temporal and spatial interaction between the carbon and water footprints of the computing systems. Our experimental characterization reveals several insights (Sec.~\ref{sec: background} and~\ref{sec:motivation}). \textit{First, we show that carbon-friendly energy sources (renewable energy sources) reduce the carbon footprint of computing systems -- which is desirable, but the availability depends on the geographical location and energy sources may also require water to generate electricity used to power the data centers (offsite water usage).} Unfortunately, some carbon-friendly energy sources (renewable sources) may need proportionally larger amounts of water to generate electricity -- \textit{creating a conflict between carbon and water co-optimization} (carbon-friendly energy sources may not be water-sustainable). Furthermore, water is also required to cool down the data centers (onsite water usage) which is significantly impacted by the weather conditions of the data center region -- \textit{introducing geographical trade-offs for water sustainability}. Motivated by these insights, we design a carbon- and water-aware job scheduler that aims to navigate these complex trade-offs in a geographically distributed data center serving parallel workloads. 

\vspace{1mm}
\noindent\textbf{Contributions.} The key contributions of \sol{} include the following highlights.
\vspace{1mm}

\noindent\textbf{Novel Carbon- and Water-Sustainability Job Scheduling Framework.} \sol{} is a novel job scheduler for parallel workloads that identifies the need and opportunity for co-optimizing carbon and water footprint for large-scale systems, and designs a simple yet effective approach to improve the sustainability of geographically distributed data centers. \sol{} reveals that allowing a small amount of delay tolerance can bring additional significant carbon- and water-footprint savings -- leveraging the characteristics that batch jobs can inherently tolerate delays~\cite{ambati2021good,zhang2022schedinspector,chen2019parties}. 
\vspace{1mm}

\noindent\textbf{\sol{} Key Ideas and Approach.} To navigate the competing carbon and water trade-offs, \sol{} designs and implements a Mixed Integer Linear Programming (MILP) based optimization -- due to MILP's inherent known advantage of configurability of objectives, low-overhead, and easier explainability~\cite{floudas2005mixed}. However, MILP methods as-is cannot operate most effectively in \sol{}'s context; to address this limitation, we augment the MILP design with the concept of delay tolerance, soft constraints, and slack management (Sec.~\ref{sec:design}). \sol{} dynamically and opportunistically exploits spatial and temporal variations in carbon and water intensity characteristics, while ensuring that jobs minimize the violations of their delay tolerance and consider inter-region movement transfer latency. Notably, \sol{} recognizes that one unit of water does not have the same value across all geographical regions (one liter of water in a water-stressed region may be more precious than in a water-abundant region~\cite{li2023making}). \sol{} systematically captures this effect of water scarcity factor~\cite{ritchie2024electricity} in its scheduling decisions -- filling a critical gap in computer system sustainability studies. 

\vspace{1mm}

\noindent\textbf{Evaluation and Open-source Framework.} \sol{} is evaluated using widely-used production traces such as Google Borg~\cite{clusterdata:Wilkes2020} and Alibaba~\cite{tian2019characterizing}, and parallel workloads from the PARSEC benchmark~\cite{zhan2017parsec3}. \sol{} provides approx. 21\% carbon-footprint and approx. 14\% water-footprint savings compared to the carbon- and water-unaware scheduler. \sol{} is available at \url{https://zenodo.org/records/14219862} -- the first open-source framework to enable exploration of carbon- and water-aware scheduling.

%% file: sections/background.tex
\section{Background}
\label{sec: background}

\subsection{Carbon Footprint in Data Center}
\label{sec:carbon_footprint}
The carbon footprint in the data center consists of two parts: operational and embodied carbon footprint~\cite{gupta2022act}.

\vspace{2mm}

\noindent\textbf{Operational Carbon Footprint.} The operational carbon footprint refers to the carbon emissions incurred while executing the workloads in the data center. The electricity consumed by the data center may come from a mix of energy sources, including both fossil fuels and renewables. The operational carbon footprint can be estimated by multiplying the energy consumption (unit kWh) and the real-time carbon intensity (unit gCO$_2$/kWh) of the energy mix used by the data center (lower is better). Carbon intensity ($\text{CO}_{2}^{\text{Intensity}}$) depends on the energy mixture of the electricity grid. 

\vspace{2mm}
\noindent\textbf{Embodied Carbon Footprint.} The embodied carbon footprint is a one-time carbon emission that occurs during server design and manufacturing. This footprint is amortized over the hardware's lifespan. Essentially, the footprint attributed to a single job is the server's embodied carbon footprint scaled by the job’s execution time over the server's lifetime.

\vspace{2mm}
\noindent\textbf{Total Carbon Footprint Estimation.} The carbon footprint for a job $j$ is the sum of its operational and embodied carbon footprint and can be expressed as:

\vspace{-3mm}
{\small
\begin{align}
\begin{split}
    \text{CO}_{2,\textit{j}} &=  \text{CO}_{2,\textit{j}}^{\text{operational}} + \text{CO}_{2,\textit{j}}^{\text{embodied}}\newline\\
    \text{CO}_{2,\textit{j}} &= E_{\textit{j}}\cdot  \text{CO}_{2}^{\text{Intensity}} + \frac{t_{j}}{T_{\text{lifetime}}}\cdot  \text{CO}_{2,\text{server}}^{\text{embodied}}
\end{split}
\end{align}}
\label{eq:carbon}
\vspace{-3mm}

Here, $\text{CO}_{2}^{\text{Intensity}}$ represents the real-time carbon intensity of the regional power grid, and it's multiplied by the energy ($E_j$) of the job to get the operational carbon. $t_j$ denotes the execution time of the job, and $T_{\text{lifetime}}$ is the overall lifetime of the server hardware. $\text{CO}_{2,\text{server}}^{\text{embodied}}$ is the total embodied carbon footprint of the server. The embodied carbon footprint for job execution is calculated by multiplying the time proportion $\frac{t_j}{T_{\text{lifetime}}}$ by the total embodied carbon footprint of the server.

\subsection{Water Footprint in Data Center}
\label{sec:back-water}
In 2023, Google published the data center's water consumption usage of 5.6 billion gallons (equivalent to irrigating 37 golf courses annually)~\cite{googlereport}, and it observed a 24\% increase compared to 2022. This sharp rise highlights the growing reliance on water-intensive cooling technologies to meet computing demands. If this trend continues without innovation in water footprint scheduling techniques, it could exacerbate water stress in regions where data centers are located, particularly in drought-prone areas. Moreover, reducing water usage in energy generation and cooling is crucial for preserving aquatic ecosystems and maintaining biodiversity. Water consumption during energy generation and cooling in the data center can disrupt natural habitats and expose aquatic life to harmful temperature fluctuations and potential pollutants. By minimizing the water footprint, we can lessen these environmental impacts, protect fish populations and other aquatic organisms, and contribute to freshwater ecosystems' overall health and sustainability.

\vspace{1mm} 

\textbf{Water Consumption Sources and Water Scarcity.} We note that both the indirect usage of water (e.g., water used during electricity generation) and the direct usage of water (e.g., water used during powering and cooling the IT infrastructure) contribute toward the overall water footprint. However, one liter of water is not equal at different geographical locations. This is captured by the metric referred to as \textbf{water scarcity}. The \emph{Water Scarcity Factor} (WSF) serves to gauge the degree of water scarcity or stress in specific regions~\cite{ritchie2024electricity}, and measure the availability of freshwater relative to demand (a higher WSF indicates that the region is more scarce). To include the impact on region-specific water scarcity, the water footprint is scaled by the WSF factor for the region to reflect the effective water footprint~\cite{ourworldindataWaterStress,wsf}. Next, we provide the basic definitions, intuitions, and estimation methodology for water footprint. 

\vspace{2mm}

\textbf{Water Footprint Categories.} For simplicity of explanation and estimation, the water footprint can be broken into two parts: \textbf{operational water footprint} and \textbf{embodied water footprint}~\cite{li2023making}. The operational water footprint can be further divided into two categories: \textit{offsite water footprint and onsite water footprint}.

\vspace{2mm}
\noindent\textbf{Operational Water Footprint (Offsite).} The offsite water footprint is the water usage resulting from using various energy sources to generate electricity for data centers -- up to 50\% of the water used by data centers can be attributed to offsite usage of water in energy generation~\cite{siddik2021environmental}. The offsite water footprint depends on the water needed by the energy source to produce electricity, and the total amount of energy generated to be supplied to the data center. 

The required amount of water needed to generate a unit amount of energy can vary depending upon the energy mix used to generate the electricity. For example, more offsite water is needed to generate 1kWh of electricity if the energy source is fossil fuel compared to a solar energy source. This dependence is captured in the \textit{energy water intensity factor} (EWIF) metric. EWIF (unit L/kWh) is dependent on the energy source being used and reflects the intensity of water consumption needed to generate the electricity. Higher EWIF reflects that the energy source is more water-thirsty. If multiple energy sources are used to generate the electricity, then, an average EWIF reflects the water consumption during electricity generation. 

The total amount of energy generated to be supplied to the data center depends on the \textit{Power Usage Effectiveness} (PUE) factor. PUE measures the energy efficiency of a data center. Ideally, PUE should be 1.0, but a higher value represents how much energy is spent on non-IT server equipment (e.g., cooling). Therefore, a job with $E_{j}$ energy uses will need a total of $\text{PUE}\times E_{j}$ energy. The following expression captures the offsite water footprint: 

\vspace{-4mm}
{\small
\begin{align}
\begin{split}
\text{H}_2\text{O}_{j}^{\text{offsite}}=\text{PUE}\times E_{j}\times\text{EWIF}\times (1+\text{WSF}_r^{\text{dc}})
\end{split}
\end{align}}
\label{eq:offsite_water}
\vspace{-4mm}

$\text{WSF}_r^{\text{dc}}$ is the water scarcity factor in the region where the data center is located and operated. This is included to capture the impact of the water scarcity factor in the region.  

\vspace{2mm}

\noindent\textbf{Operational Water Footprint (Onsite).} Besides water consumption during the electricity generation phase, the data centers directly use water onsite for cooling purposes~\cite{ebrahimi2014review}. The onsite water footprint refers to the amount of water evaporated and dissipated during heat transfer and blowdown in the cooling process. The onsite water footprint can be estimated by multiplying the energy consumed during job execution ($E_j$) by the \emph{Water Usage Effectiveness} (WUE). The WUE metric quantifies the water required to dissipate heat per unit of energy generated, measured in L/kWh (lower is better). WUE directly depends on the wet bulb temperature specific to the data center's location and is influenced by temporal temperature fluctuations within a region and across regions. This is why two data centers with the same energy-efficiency (PUE) value can have very different onsite water footprints due to differences in the region they are located and the weather in the region over time. The following expression captures the onsite water footprint: 

\vspace{-4mm}
{\small
\begin{align}
\begin{split}
\text{H}_2\text{O}_{j}^{\text{onsite}}=E_{j}\times\text{WUE}\times (1+\text{WSF}_r^{\text{dc}})
\end{split}
\end{align}}
\label{eq:onsite_water}
\vspace{-4mm}

$\text{WSF}_r^{\text{dc}}$ is the water scarcity factor in the region where the data center is located and operated. This is included to capture the impact of the water scarcity factor in the region. 
 
\vspace{2mm}

\noindent\textbf{Embodied Water Footprint.} Similar to the embodied carbon footprint, the embodied water footprint is the water consumption generated during server design and manufacturing. Because of the lack of public data on embodied carbon, we use an alternative approach to estimate it. We start by collecting the total embodied carbon footprint of the server~\cite{davy2021building,auger2021open}, and then, multiply it by the carbon intensity of the server's manufacturing place to estimate the total energy consumption during design and manufacturing. Then, this total energy consumption during design and manufacturing can be used to estimate the water consumption during the design and manufacturing -- by multiplying the Energy Water Intensity Factor (EWIF) by the total energy consumption during manufacturing ($E_{\text{manufacturing}}$), as expressed below: 

\vspace{-1em}
{\small
\begin{align}
\begin{split}
\text{H}_2\text{O}_{\text{server}}^{\text{embodied}}=E_{\text{manufacturing}}\times\text{EWIF}\times(1+ \text{WSF}_r^{\text{server}})
\end{split}
\end{align}}
\label{eq:embodied_water}
\vspace{-1em}

$\text{WSF}_r^{\text{server}}$ is the water scarcity factor in the region where the server is manufactured. 

\vspace{2mm}
\noindent\textbf{Total Water Footprint.} The total water footprint of a job can be expressed as:

\vspace{-1em}
{\small
\begin{align}
\begin{split}
    \text{H}_2\text{O}_{\textit{j}} =   \text{H}_2\text{O}_{\textit{j}}^{\text{operational}} +  \text{H}_2\text{O}_{\textit{j}}^{\text{embodied}}\newline\\
     \text{H}_2\text{O}_{\textit{j}} = \text{PUE}\times E_{j}\times\text{EWIF}\times (1+\text{WSF}_r^{\text{dc}}) \\
     +E_{j}\times\text{WUE}\times (1+\text{WSF}_r^{\text{dc}})+
     \frac{t_{j}}{T_{\text{lifetime}}}\cdot  \text{H}_{2}\text{O}_{\text{server}}^{\text{embodied}}
\end{split}
\end{align}
\label{eq:total_water_footprint}}
\vspace{-0.8em}

\noindent\textbf{Water Intensity.} \textit{To make the interpretation of results and analysis easier, we propose the concept of water intensity.} Similar to carbon intensity, water intensity captures the spatial and temporal variation in water consumption and the stress it causes on the region (where the data center is located). Intuitively, water intensity metric should be higher (lower is desirable -- similar to carbon intensity, where lower is better) if (a) the water scarcity is high (more water stress), (b) water usage effectiveness is high (more water required to dissipate heat per unit of energy generated in the data center, due to unfavorable weather condition), and (c) high PUE and EWIF (i.e., more water is needed - hence, water intensity is higher). Thus, water intensity can be simply captured by focusing on these components in the onsite and offsite water footprint. This can be expressed as:

\vspace{-1em}
{\small
\begin{align}
\begin{split}
\text{H}_2\text{O}^{\text{Intensity}} = (\text{WUE}+\text{PUE}\cdot\text{EWIF})\cdot(1+\text{WSF}_r^{\text{dc}})
\end{split}
\label{eq:water_intensity}
\end{align}
}
\vspace{-1em}

Note that embodied water footprint is not a part of the water intensity expression because embodied water footprint is relevant only to the region where the server is manufactured, but water intensity is focused on water requirements in the data center region. With this background, next, we discuss the motivations behind \sol{}'s design. 

%% file: sections/motivations.tex
\section{Experimental Insights and Motivation }
\label{sec:motivation}

\noindent\textbf{Observation 1}: \emph{The amount of water required to generate electricity for powering data centers from different energy sources (renewable energy sources vs. fossil energy sources) is significant and varies across energy sources. The water consumption to generate energy from carbon-friendly and carbon-intensive energy sources (resulting in the EWIF factor) varies significantly. In general, carbon-friendly energy sources can have higher water footprint requirements -- highlighting the tension between carbon and water as sustainability goals.}

\begin{figure}[t]
    \centering
\includegraphics[scale=0.48]{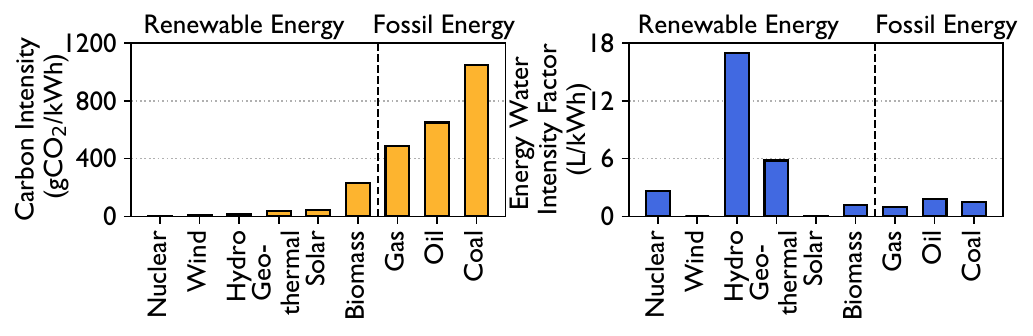}
    \vspace{-4mm}
    \caption{Carbon intensity and water requirements (Energy Water Intensity Factor (EWIF)) for energy generation for different types of energy sources. Carbon-friendly energy sources can have higher water needs for energy generation (EWIF), contributing to a higher offsite water footprint.}
    
    \label{fig:motiv1}
    \vspace{-4mm}
\end{figure}

\begin{figure*}[t]
    \centering
    \includegraphics[scale=0.46]{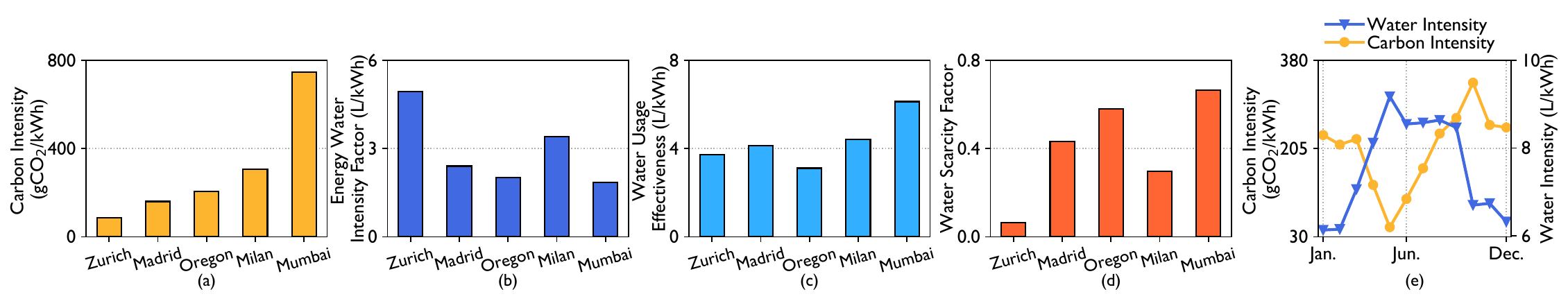}
    \vspace{-3mm}
    \caption{Carbon Intensity, EWIF, WUE, WSF varies across different geographical regions (average values for the year 2023, region labels are sorted according to the carbon intensity). Carbon intensity and water intensity show temporal variations.}
    \label{fig:motiv2}
    \vspace{-4mm}
\end{figure*}

Fig.~\ref{fig:motiv1} characterizes the notable quantitative differences
in carbon intensity among various energy sources and their corresponding EWIF (counts toward offsite water footprint) for power generation. We show that different types of energy sources exhibit varying carbon intensities due to the differing amounts of carbon they produce during power generation~\cite{bruckner2014technology}. Typically, carbon-intensive energy sources (fossil fuels) have higher carbon intensities compared to carbon-friendly energy sources (renewable energy). For example, coal has a carbon intensity of 1050 gCO$_2$/kWh is roughly 62$\times$ higher than the carbon intensity of hydropower, which is 17 gCO$_2$/kWh. Interestingly, while renewable energy sources tend to have a lower carbon intensity, they may exhibit higher EWIFs, leading to a greater offsite water footprint. For example, hydropower, despite its low carbon intensity, has a high EWIF of 17L/kWh, which is 11$\times$ greater than that of coal. Other renewable energy sources require a significant amount of water for cooling and irrigation, leading to higher EWIF compared to fossil fuels. Other renewable energy sources like biomass require significant water for growing feedstock, which is then converted into energy, and for cooling, leading to higher EWIF compared to fossil fuels. Thus, depending on the energy source mix used by power grids, the carbon and water footprint can vary. Additionally, carbon-efficient energy sources can be water-inefficient, and vice-versa.

\vspace{2mm}
\noindent\textbf{Observation 2}: \emph{Geographical regions with availability to low carbon intensity energy sources (using more carbon-friendly energy sources for power generation) can be water-stressed regions -- due to high EWIF (high water requirements to generate electricity) and high WUE (unfavorable weather conditions requiring more water for cooling). To make it worse,  these carbon-friendly but water-stressed regions may have a relatively higher water scarcity factor. Consideration of the water scarcity factor changes the trade-off between carbon and sustainability (carbon vs water intensity).}

\vspace{2mm}

From Fig.~\ref{fig:motiv2}, we observe the average values of different factors impacting the carbon and water footprint of data centers located in different geographical regions. The carbon intensity and the water required to generate electricity to provide energy to a data center (EWIF) varies significantly among different regions due to the difference in the energy sources used by power grids. For example,  Zurich (Switzerland) has low carbon intensity (Fig.~\ref{fig:motiv2}(a)) as the power grid in Zurich uses renewable energy sources including biomass and hydropower~\cite{ritchie2024electricity}. However, since these sources require a high amount of water for generation, the EWIF of Zurich is the highest (Fig.~\ref{fig:motiv2}(b)), despite having the lowest carbon intensity.  Mumbai (India) has a very high carbon intensity as power grids in Mumbai use a large share of non-renewable sources  (e.g., coal and oil) to produce electricity. However, since these energy sources have low EWIF (Fig.~\ref{fig:motiv1}), the regional EWIF is relatively low (Fig.~\ref{fig:motiv2}(b)). 

While EWIF impacts the offsite water footprint, the water usage effectiveness (WUE) impacts the onsite water footprint (Sec.~\ref{sec:back-water}), required for cooling a data center. Specifically, WUE depends on the wet bulb temperature of a region, which is impacted by the regional temperature and relative humidity. The resulting region-wise WUE variation (Fig.~\ref{fig:motiv2}(c)) is different from the variations of carbon intensity and EWIF. The difference in region-wise variations of these factors makes it challenging to jointly optimize for carbon and water footprint by distributed job scheduling across multiple regions.

Additionally, from Fig.~\ref{fig:motiv2}(d), we observe that geographic regions that rely on carbon-friendly energy sources are often regions experiencing high water stress (resulting in high water scarcity factors) -- for example, Madrid (Spain).  Additionally, certain regions can have low EWIF (power grids use energy sources that consume less water for energy production), but have a high water scarcity (e.g., Mumbai (India) and Oregon (USA)). Thus, in spite of power grids using water-efficient energy sources, it can be unwise to schedule a significant number of jobs in data centers located in water-scarce regions, as it can severely impact the water availability in such regions for other purposes. Combining the water scarcity factor with the water footprint, along with the variations in carbon intensity, makes distributed job scheduling challenging. 

 From Fig.~\ref{fig:motiv2}(a)-(d), we observe that the factors impacting the water intensity (EWIF, WUE, and WSF) and carbon intensity vary spatially across regions. However, as the energy mix used by power grids in a region changes with time, they also have temporal variations. This leads to temporal variations in carbon and water intensity, as shown in Fig.~\ref{fig:motiv2}(e) for Oregon in 2023. Interestingly, we note that certain periods can have high carbon intensity with low water intensity and vice versa. Thus, there is an opportunity to leverage this temporal variation of carbon and water intensity.

\vspace{2mm}
\noindent\textbf{Observation 3}: \emph{Carbon and water footprint optimization can be performed via effective job scheduling across different data center regions, leveraging temporal variations in carbon and water intensity. The competing nature of carbon and water optimization makes the optimization challenging, but allowing jobs to tolerate some delay increases the opportunity scope.}
\vspace{2mm}

To analyze the benefits we observe via distributed scheduling of jobs among different data center regions leveraging temporal variations of carbon intensity and water intensity, we design greedy-optimal solutions: Carbon-Greedy-Optimal, and Water-Greedy-Optimal (referred to as Carbon-Greedy-Opt, and Water-Greedy-Opt and discussed in Sec.~\ref{sec:methodology}). 

\begin{figure}[!t]
    \centering
\includegraphics[scale=0.49]{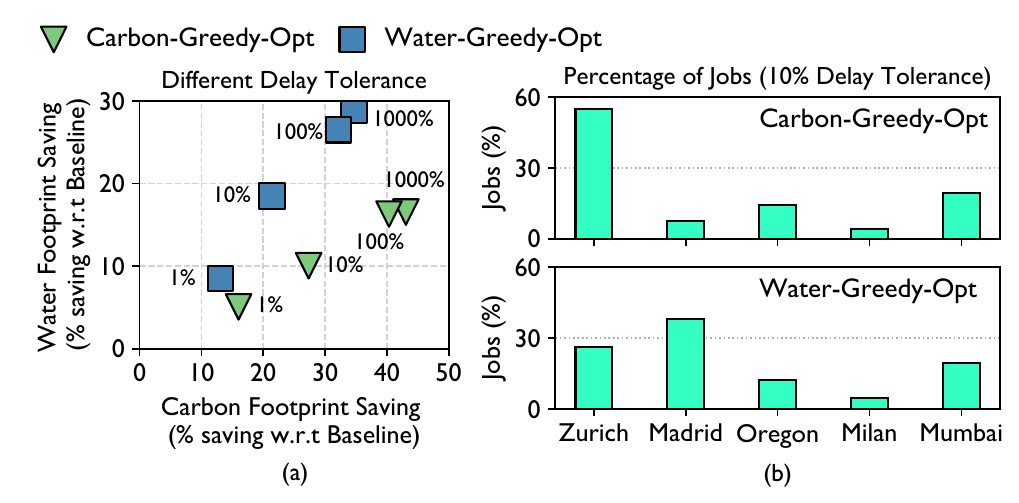}
    \vspace{-7mm}
    \caption{Quantifying optimal solution benefits, opportunity scope due to delay tolerance, and impact on job distribution across regions.}
    \label{fig:motiv4}
    \vspace{-5mm}
\end{figure}

These optimal solutions are infeasible in practice, as they require knowledge about a job's execution time and future variations of carbon and water intensities.  We execute jobs following Google Borg trace in five data center regions across the world (details about jobs and execution based on trace in Sec.~\ref{sec:methodology}). We observe from Fig.~\ref{fig:motiv4}(a) that the Carbon-Greedy-Opt solution performs sub-optimally in terms of water footprint and the Water-Greedy-Opt solution is suboptimal in terms of carbon footprint -- demonstrating the challenging nature of carbon and water footprint co-optimization. This happens due to the spatio-temporal mismatch of the factors affecting the carbon and the water footprint. 

Furthermore, Fig.~\ref{fig:motiv4}(a) also reveals that if jobs could tolerate some latency in their service time -- the opportunity for carbon and water footprint savings improves. This is because delay tolerance allows the scheduler to transfer jobs across regions to opportunistically leverage lower carbon and water intensity. \textit{We define delay tolerance as an allowable \% increase in the service time of a job compared to its execution time of the job if it had zero transfer latency and queuing delay}. For example, a 25\% delay tolerance indicates the scheduler has the flexibility to delay its execution such that the total service time remains within 1.25$\times$ of the execution time if the job was scheduled right away. 

To understand the impact of delay tolerance, from Fig.~\ref{fig:motiv4}(a) we also observe that if delay increases, the carbon and water footprint savings improve even further. When jobs can tolerate higher delay, they can explore more choices of getting transferred to different regions (hiding service time-delay effects). More transfer options to different regions also mean more opportunity to leverage different values of carbon intensity and water intensity, thus resulting in potentially higher carbon and water footprint savings. However, as expected, the returns are diminishing as we increase the tolerance from 10\% to 1000\%.

Next, in Fig.~\ref{fig:motiv4}(b) we observe the percentage distribution of jobs scheduled across the five data center regions following the Carbon-Greedy-Opt and Water-Greedy-Opt strategies, for a 10\% delay tolerance. We make two observations. First, as expected, the jobs are distributed across all regions. Even though some regions may be naturally better suited for achieving a carbon- or water-optimal solution, due to temporal variations in carbon and water intensity, jobs are distributed across all regions. This means \textit{no single particular region} is always the optimal choice in terms of minimizing carbon or water footprint. Second, the distributions of jobs across the different regions for carbon and water-optimal solutions are significantly different. 

%% file: sections/design.tex
\section{\sol{} Design}
\label{sec:design}

\noindent\textbf{\sol{} Objective and Desirable Design Goals.} The primary objective of \sol{} is to minimize the carbon footprint and water footprint when executing jobs in geographically distributed data centers, under different specified levels of delay tolerance. The system design of \sol{} should strive to meet several desirable requirements. In particular, the core scheduler engine of \sol{} should be configurable to balance different objectives, the decisions should be interpretable, the decision-making overhead should be low, and the delay tolerance limit should not be violated for jobs. While the scheduler cannot have futuristic information (e.g., future carbon and water intensity for different regions), it should co-optimize jobs that are invoked together or nearby in time based on the current carbon and water intensity of all available regions.

\begin{figure}[t]
    \centering
    \includegraphics[scale=0.36]{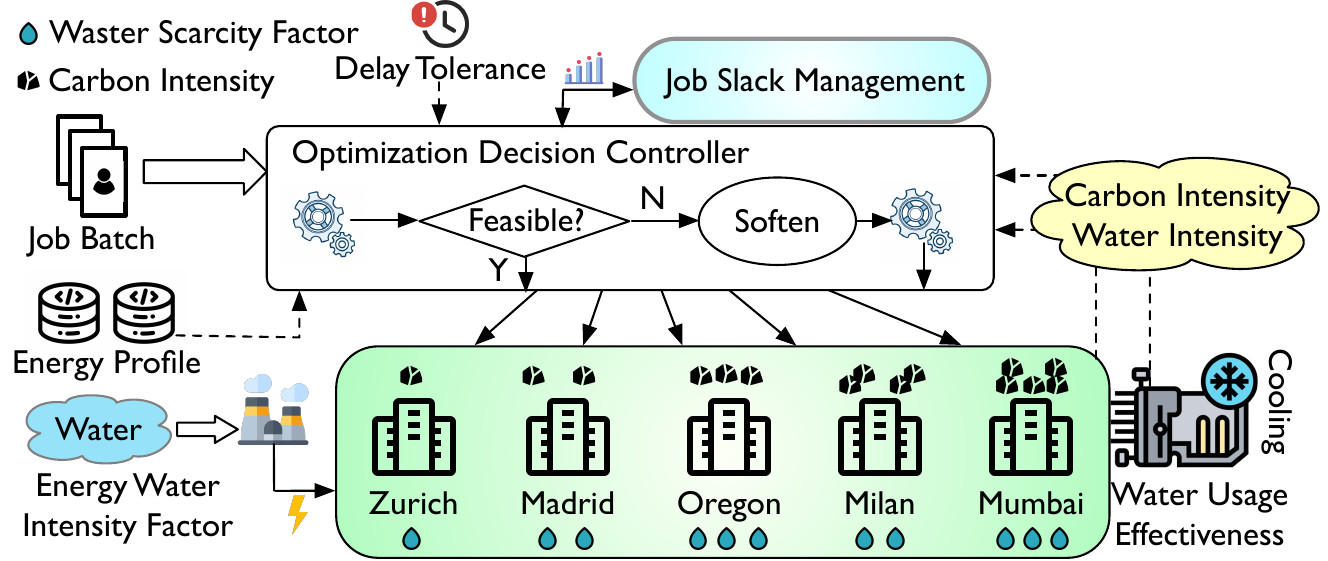}
    \caption{Design overview of \sol{}.}
    \label{fig:system}
        \vspace{-5.5mm}
\end{figure}

\vspace{2mm}
\noindent\textbf{System Overview of \sol{}.} \sol{} collects and holds all detailed information about the incoming jobs, including job dependencies, execution files, and metadata. Each job specifies a home region, indicating where the users have submitted their jobs. If \sol{} determines to move the job away from its home region, it considers the transfer latency impact on the service time and allowable delay tolerance. \sol{} Optimization Decision Controller decides where to execute which job while considering the current carbon and water intensity of all regions and collected current mean estimates about job execution time and energy from their previous executions; however, these estimates can be inaccurate. 

Recall that, as desirable properties, the core scheduler engine of \sol{} should be configurable to balance different objectives, the decisions should be interpretable, and the decision-making overhead should be low. Therefore, \sol{} employs a Mixed Integer Linear Programming (MILP) based approach for scheduler optimization. While other approaches are certainly viable (e.g., machine learning) and potentially effective, \sol{} chose MILP to keep the overhead practical (as confirmed in evaluation) and decisions interpretable/scheduling objectives configurable, as discussed shortly. \sol{}'s Optimization Decision Controller dispatches jobs to selected regions and reconfigures the job queue to ensure that jobs with the least slack in meeting their delay tolerance are prioritized. The overview of \sol{} is shown in Fig.~\ref{fig:system}.

\vspace{2mm}
\noindent\textbf{\sol{} Objective Function.}
Recall that carbon and water are often conflicting goals both in time and space, and \sol{}'s goal is to co-optimize both the carbon and water footprint considering all the regions and all available jobs ready to be executed. Formally, the objective function can be expressed as:

\vspace{-3mm}
{\small
\begin{align}
\label{eq:obj_func}
f(j,r)=\lambda_{\text{CO}_2}\frac{\text{CO}_{2}(j,r)}{\text{CO}_{2,j}^{\text{max}}} + \lambda_{\text{H}_2\text{O}}\frac{\text{H}_2\text{O}(j,r)}{\text{H}_2\text{O}_{j}^{\text{max}}}
\end{align}}
\vspace{-3mm}

Here, $f(j,r)$ is the optimization objective function, $j$ is the job to be scheduled and $r$ is the region for execution selected from the available regions. $\lambda_{\text{CO}_2}$ and $\lambda_{\text{H}_2\text{O}}$ are configurable parameters that the cloud service provider can set to control the optimization degree of carbon and water footprint, respectively. The sum of these parameters equals one. $\text{CO}_2(j,r)$ and $\text{H}_2\text{O}(j,r)$ represent the carbon footprint and water footprint for executing the job $j$ in the region $r$. ${\text{CO}_{2,j}^{\text{max}}}$ and ${\text{H}_2\text{O}_{j}^{\text{max}}}$ are the maximum carbon and water footprint of job $j$ executed in the region with the highest carbon intensity and water intensity, respectively. \sol{} leverages these maximum values for normalization to ensure that one objective does not skew the optimization. For simplicity, we assume that data centers in different regions use the same generation of hardware with no performance differences, while transfer latency between regions is considered in the optimization process as described below.

\vspace{2mm}
\noindent\textbf{Problem Formulation.} We now formulate this job scheduling problem using Mixed Integer Linear Programming (MILP). The inputs of \sol{} are $M$ jobs ready for execution, and data centers are distributed across $N$ different regions. \sol{} uses one optimization variable, denoted as $x_{m,n}$. Here, $m$ and $n$ indicate the $m$-th job and $n$-th region, and $x_{m,n}$ is a Boolean variable, where $x_{m,n} = 1$ indicates that the $m$-th job is assigned to the data center in the region $n$ for execution, and $x_{m,n} = 0$ indicates otherwise. Our goal is to minimize the objective function defined in Eq.~\ref{eq:obj_func}, as $f(m,n)$. This problem can be formulated as:

 \vspace{-3mm}
{\small
\begin{align}
\label{eqal:obj}
\begin{split}
     \operatorname*{min}\sum_{m=1}^{M}\sum_{n=1}^{N}x_{m,n}\cdot[\lambda_{\text{CO}_2}\frac{\text{CO}_{2}(m,n)}{\text{CO}_{2,j}^{\text{max}}} +\lambda_{\text{H}_2\text{O}}\frac{\text{H}_2\text{O}(m,n)}{\text{H}_2\text{O}_{j}^{\text{max}}}\\
     +\lambda_{\text{ref}}\cdot(\lambda_{\text{CO}_2}\cdot \text{CO}_{2,n}^{\text{ref}} +\lambda_{\text{H}_2\text{O}}\cdot \text{H}_2\text{O}^{\text{ref}}_{n} )]
\end{split}
\end{align}}

\sol{} leverages the historical carbon footprint and water footprint (normalized) of every region in a time window as the history learner to the objective function ($\text{CO}_{2,n}^{\text{ref}}$ and $\text{H}_2\text{O}^{\text{ref}}_{n}$ ). \sol{} utilizes configurable parameters ($\lambda_{\text{ref}}$) to control the influence of history on the optimization process. 

The optimization variable ($x_{m,n}$) must satisfy two constraints. First, \sol{} ensures that each job is assigned to only one data center in a specific region for execution only if the capacity is available (Eq.~\ref{con1}). Second, each region has a limited capacity. The number of jobs assigned to any region cannot exceed its capacity (Eq.~\ref{con2}). We use $\text{cap}(n)$ denoting the remaining capacity of the data centers in region $n$. The remaining capacity is gathered using the monitor in \sol{}. We formalize these 2 constraints as the following:

\vspace{-3mm}
{\small
\begin{align}
\label{con1}
\begin{split}
s.t. \quad\sum_{n=1}^{N} x_{m,n} = 1, \quad \forall m \in \{1, 2, \ldots, M\}
\end{split}
\end{align}}
\vspace{-3mm}

\vspace{-3mm}
{\small
\begin{align}
\label{con2}
\begin{split}
s.t. \quad\sum_{m=1}^{M} x_{m,n} \leq \text{cap}(n), \quad \forall n \in \{1, 2, \ldots, N\}
\end{split}
\end{align}}
\vspace{-3mm}

These two constraints must be satisfied. If the capacity is not enough among all regions combined, \sol{} uses its slack management to prioritize certain jobs to avoid violating delay tolerance. 

\vspace{2mm}
\noindent\textbf{Delay Tolerance Constraint.} Recall that delay tolerance, a configurable parameter for the service provider, determines the maximum allowable service time for a job (relative to its pure execution time in its home region - that is, no queuing or transfer latency). The execution time of a single job is gathered from the database, and the maximum service time allowance is calculated based on this execution time. Delay tolerance (denoted as TOL\%) specifies the percentage by which the total service time (including transfer latency and queuing delay) can exceed the execution time. The delay constraint can be expressed as:

\vspace{-3mm}
{\small
\begin{align}
\label{con3}
\begin{split}
s.t. \quad \sum_{n=1}^{N} x_{m,n}\cdot\frac{L_{m,n}}{t_{m,n}} \leq(\text{TOL} \%),\ \ \forall m \in \{1, 2, \ldots, M\}
\end{split}
\end{align}}
\vspace{-3mm}

$L_{m,n}$ is the transfer latency when transferring $m$-th job from its original region to $n$-th region. $t_{m,n}$ denotes the execution time of $m$-th job in $n$-th region. Note that, different regions utilize the same type of hardware, so there are not many performance variations. Higher delay tolerance allows \sol{} to achieve greater carbon and water savings.

\vspace{2mm}
\noindent\textbf{MILP Optimization.} The problem formulation is Eq.~\ref{eqal:obj} with the constraints (Eq.~\ref{con1}, ~\ref{con2}, ~\ref{con3}). \sol{} solves the MILP problem with PuLP~\cite{mitchell2011pulp} to determine the global optimal scheduling decisions of all jobs. PuLP is a Python package integrating multiple algorithms (cutting plane, branch-and-bound) for solving the MILP problem. Its optimization results are equally accurate and the optimization process is more configurable compared with the heuristic algorithms (genetic algorithm, anneal simulated algorithm), also unlike machine learning models, where the reasoning behind outcomes might be opaque, MILP outcomes can be traced directly back to the mathematical model. Details of the solver and decision-making overhead are in Sec.~\ref{sec:methodology} and Sec.~\ref{sec:evaluation}. However, MILP solver has two challenges in \sol{}'s context: (a). \emph{Fragility under hard constraints}: MILP solver can fail to provide a solution if all the constraints cannot be satisfied, making it less flexible in certain situations. (b). \emph{Statelessness w.r.t. delay tolerance violation}: MILP problem itself is stateless, it does not retain information about which jobs may be closer to delay tolerance. The solver only solves the problem given the required parameters, it doesn't have the awareness of the potential for delay tolerance violation if the job arrived in the past and has been delayed in anticipation of better carbon and water savings. To address these issues, we show two extensions next in the Decision Controller to make \sol{} effective and robust and meet the design properties.

\vspace{2mm}
\noindent\textbf{\sol{}'s Soft Constraints.} When the solver cannot provide feasible and robust optimization outputs, \sol{} softens the delay tolerance constraint (Eq.~\ref{con3}) based on the penalty methods~\cite{boyd2004convex}. The penalty method is an optimization technique used to solve constrained problems by converting them into unconstrained ones. A penalty term is added to the objective problem, which imposes a cost for violating the constraints. The magnitude of the penalty increases as the solution moves further from satisfying the constraints, thereby guiding the optimization process toward near-optimal solutions. By adjusting the penalty value, the delay penalty can balance between optimizing the objective and adhering to the constraints, making \sol{} effective.

For more details, \sol{} converts the delay tolerance constraint (Eq.~\ref{con3}) into a new relaxed constraint (Eq.~\ref{penalty_con}). This relaxation is achieved by introducing relaxed penalty variables that account for potential violations of the original delay tolerance. The penalty associated with these variables is then incorporated into the objective problem. The modified optimization problem can now be expressed as follows:

 \vspace{-5mm}
{\small
\begin{align}
\label{eqal:penalty_prob}
\begin{split}
     \operatorname*{min}\sum_{m=1}^{M}\sum_{n=1}^{N}x_{m,n}\cdot[\lambda_{CO_2}\frac{CO_{2}(m,n)}{CO_{2,j}^{\text{max}}} +\lambda_{H_{2}O}\frac{H_{2}O(m,n)}{H_{2}O_{j}^{\text{max}}}\\
     +\lambda_{ref}\cdot(\lambda_{CO_2}\cdot CO_{2,n}^{ref} +\lambda_{H_{2}O}\cdot H_2O^{ref}_{n} )]+ \sigma\cdot\sum_{m=1}^{M}\sum_{n=1}^{N}P_{m,n}
\end{split}
\end{align}}
\vspace{-6mm}

\vspace{-3mm}
{\small
\begin{align}
\label{penalty_con}
\begin{split}
s.t. \quad \sum_{n=1}^{N} x_{m,n}\cdot\frac{L_{m,n}}{t_{m,n}} \leq(\text{TOL} \%+P_{m,n}),\ \ \forall m \in \{1, 2, \ldots, M\}
\end{split}
\end{align}}
\vspace{-4mm}

\noindent\textbf{Job Slack Management.} \sol{} designs slack management to determine which jobs are closer to their respective delay tolerance violation (recall that MILP does not retain state information across interactions). \sol{}'s slack manager determines which jobs should be prioritized for scheduling and where they should be executed. Once these prioritized jobs are scheduled, any remaining unscheduled jobs will be considered in the next iteration. 

More formally, when the number of jobs exceeds the total available capacity \(\sum_{n=1}^{N}\text{cap}(n)\), meaning the current cluster cannot handle this many jobs, \sol{} composes the following urgency score to determine the priority of each job:

\vspace{-3mm}
{\small
\begin{align}
\label{pri_score}
\begin{split}
\text{Urgency} = \text{TOL}\%\cdot t_m -L_{m}^{\text{avg}}-(T_{m}^{\text{start}} - T^{\text{current}})
\end{split}
\end{align}}
\vspace{-3mm}

\begin{algorithm}[t]
\small{
\caption{\sol{}'s Scheduling Framework}
\label{alg1}
    \begin{algorithmic}[1]
        \State \textbf{Input:} Job Batch: $J = \{j_m |m \in [1,M]\}$.
        \State \textbf{Output:} Decisions $D$.
        \State Get the all jobs that need to be scheduled 
        $J_\textit{all}=J\cup J_{\text{delay}} $.
        \State Get energy ($E_{j})$, execution time ($t_{j}$), latency, capacity ($\text{cap}(n)$).
        \If{M$>\sum_{n}\text{cap}(n)$}
        \State Use \sol{} slack manager to pick top $\sum_{n}\text{cap}(n)$ jobs.
        \State Use the \sol{} soft constraint enabled Decision Controller: $D$.
        \Else 
        \State Use the \sol{} hard constraint enabled Decision Controller: $D$.
        \If{D is not a feasible solution}
          \State Use the \sol{} soft constraint enabled Decision Controller: $D$.
         \EndIf 
         \EndIf
    \end{algorithmic}
    }
\end{algorithm}

\sol{} ranks all the jobs based on the urgency score from smaller to higher. To meet the delay tolerance constraint, \sol{} prioritizes the jobs according to the urgency score. $\text{TOL}\%\cdot t_m$ is the maximum allowable service time for serving $m$-th job. $L_{m}^{\textit{avg}}$ is transfer latency across all available regions. $T_{m}^{\textit{start}}$ is the time when the controller receives the $m$-th job, and $T^{\textit{current}}$ is the current time when calculating the urgency score. The last component $(T_{m}^{\textit{start}} - T^{\textit{current}})$ illustrates how long the job has been waiting.

\textbf{Summary.} The scheduling framework is summarized in Algorithm~\ref{alg1}. Next, we provide the evaluation of \sol{}'s effectiveness and robustness.

%% file: sections/methodology.tex
\section{Experiment Methodology}
\label{sec:methodology}

\noindent\textbf{Experimental Setup.} 
All experimental measurements are performed on a 175-node cluster containing machines equally distributed across five different AWS regions: \texttt{eu-central-2} (Zurich), \texttt{us-west-2} (Oregon), \texttt{eu-south-2} (Spain), \texttt{eu-south-1} (Milan), \texttt{ap-south-1} (Mumbai). The servers are AWS \texttt{m5.metal} general purpose bare-metal machines, each equipped with four 24-core, 384 GiB memory Intel Xeon Platinum 8175 CPUs with 25 Gibps network. The execution files and dependencies of jobs are compressed into \texttt{.tar}, and transferred across regions using \texttt{SCP}. 

The \sol{}'s Decision Controller is driven by a module leveraging PuLP~\cite{mitchell2011pulp} - a Python library for MILP, offering a high-level configurable interface to define optimization problems. The optimization objective is solved using various GLPK solver, which adopts the cutting plane and branch-and-bound algorithms. In our default setting, \sol{} assigns equal weight (0.5) to both $\lambda_{\text{CO}_2}$ and $\lambda_{\text{H}_2\text{O}}$ to optimize carbon and water footprint, and a weight of 0.1 to $\lambda_{\text{ref}}$ in the History Learner with a window size of 10; these parameters are configurable and were not hand-tuned to achieve better performance. The delay tolerance ranges from 25\% to 100\%, and we set the PUE to 1.2~\cite{shehabi2016united}. Our full experimental framework including the experimental measurements, experimental methodology, codebase for trace-simulation, and experiments will be open-sourced.

\begin{table}
\centering
  \vspace{-3mm}
\caption{Benchmark workloads used in \sol{}.}
\vspace{-2mm}
\scalebox{0.7}{
\begin{tabular}{cc}
\toprule
\textbf{Name} & \textbf{Benchmarks} \\ 
\midrule
\midrule
\textbf{PARSEC} &\makecell[c]{Dedup, Netdedup (Data Compression); Canneal (Engineering);\\Blackscholes, Swaptions (Financial Analysis)}\\
\midrule
\textbf{CloudSuite} &  \makecell[c]{Data Caching; Graph Analytics; Web Serving;\\ Memory Analytics; Media Streaming}\\
\bottomrule
\end{tabular}}
\vspace{-5mm}
\label{table:method2}
\end{table}

\vspace{2mm}
\noindent\textbf{Evaluated Workloads and Traces.} \sol{} is evaluated using a wide range of ten benchmarks from the CloudSuite benchmark suite~\cite{ferdman2012clearing} and the PARSEC-3.0 benchmark suite~\cite{zhan2017parsec3}. The selected benchmarks represent a wide range of scientific domains, such as graph analysis, machine learning, parallel computing, etc (Table~\ref{table:method2}). When executing jobs, we ensure that job inter-arrival follows the industry-grade traces from Google (Borg cluster traces~\cite{clusterdata:Wilkes2020}) to capture the characteristics observed on production-scale cloud computing platforms. Our evaluated trace contains ten-day data from the Google trace, including more than 230000 jobs. In the evaluation (Sec.~\ref{sec:evaluation}), we evaluate \sol{} with another widely-used cloud trace, Alibaba VM Cloud trace~\cite{tian2019characterizing} to further demonstrate the effectiveness of \sol{}. The average utilization of the evaluated cluster setup is approx. 15\% (based on the Borg cluster traces and number of servers used in the setup) -- aligned well with widely-reported data center utilization trends~\cite{cheng2018analyzing,leverich2014reconciling}, but we also perform sensitivity study to demonstrate \sol{}'s at both higher (25\%) and lower (5\%) utilization levels in Sec.~\ref{sec:evaluation}.

\vspace{2mm}
\noindent\textbf{Carbon, Water Footprint Estimation.}
Carbon and water footprint estimation follows the formulas outlined in Sec.~\ref{sec: background}. Energy consumption is measured using the Likwid~\cite{psti} to read the RAPL~\cite{khan2018rapl} energy information. \sol{} utilize a publicly available dataset~\cite{Davy_2024} to determine the total embodied carbon footprint of the \texttt{m5.metal} instance. Additionally, \sol{} uses carbon intensity data from July 2023 across five different regions, sourced from
Electricity Maps~\cite{electricitymap}.

The Energy Water Intensity Factor (EWIF) of each energy source is obtained from a widely-used open-source dataset source~\cite{macknick2011review, macknick2012operational}. The EWIF of energy sources influences the water intensity estimation; therefore, we also conduct a robustness analysis in Sec.~\ref{sec:evaluation}. Electricity Maps is used to feed the real-time breakdown information of energy mix for each region. We also evaluate the effectiveness of \sol{} when the recent data~\cite{reig2020guidance} from World Resources Institute is used. Based on the energy mix breakdown and the EWIF of individual energy sources, \sol{} calculates the regional EWIF for estimating the offsite water footprint. For onsite water footprint, \sol{} adopts the web bulb temperate of a data center's region from Meteologix~\cite{meteologix} to determine the Water Usage Effectiveness (WUE)~\cite{li2023making}. The Water Stress Factor (WSF) of different data center regions is collected from an open-source dataset~\cite{ourworldindataWaterStress}.

\vspace{2mm}

\noindent\textbf{Relevant Techniques.} We compare \sol{} with the following scheduling schemes:

\vspace{1mm}
\noindent\textbf{Baseline.} The baseline scheme contains a vanilla scheduling framework, every job is executed in its home region. This scheme does not explore the potential of carbon and water savings via migration to other locations or opportunistically delaying the execution at home locations in anticipation of carbon or water savings.

\vspace{1mm}
\noindent\textbf{Carbon-Greedy-Opt, Water-Greedy-Opt.} Both schemes are two infeasible schemes that only optimize either carbon or water footprint, respectively. Both schemes have the future knowledge of carbon intensity and water intensity for each region, jobs can be scheduled later to get more carbon and water savings. Note that, the maximum delay is bound by the delay tolerance constraint. Both schemes can have varied carbon and water footprints due to different delay tolerances. We use the brute force method to find the overall carbon and water footprint.
These methods are not truly optimal since they make the scheduling decision without knowing the characteristics of future job arrivals.

\vspace{1mm}

\noindent\textbf{Round-Robin and Least-Load Techniques.} Round-Robin and Least-Load are two common load-balancing algorithms used to distribute incoming parallel jobs across multiple servers to optimize resource utilization and ensure high availability. The Round-Robin works by sequentially distributing jobs to each server across regions, one after the other, in a circular order. The Least-Load algorithm dynamically distributes parallel jobs based on the current load of each region. It sends the job to the region with the lowest resource usage, ensuring a more balanced distribution. Both of these algorithms are not aware of the carbon and water savings.

\vspace{1mm}

\noindent\textbf{Ecovisor.} Ecovisor is a state-of-the-art virtualization container designed to execute applications with carbon efficiency~\cite{souza2023ecovisor}. Ecovisor is a novel technique that uses the carbon scaler to control the power efficiency of the jobs. Ecovisor operates by leveraging solar energy, grid power, and a virtual battery to manage the power usage of the container. We note that Ecoviser only targets the carbon footprint metric (but not the water footprint), so the scope of \sol{} and Ecovisor are fundamentally different. Nevertheless, we include Ecovisor to provide a more comprehensive evaluation to the readers. We use a customized implementation of Ecovisor~\cite{souza2023ecovisor} to enhance its functionality and use the setting of \sol{}. Refer to Sec.~\ref{sec:evaluation} for more details about the comparison with \sol{}.

\begin{figure}[!t]
    \centering
    \includegraphics[scale=0.49]{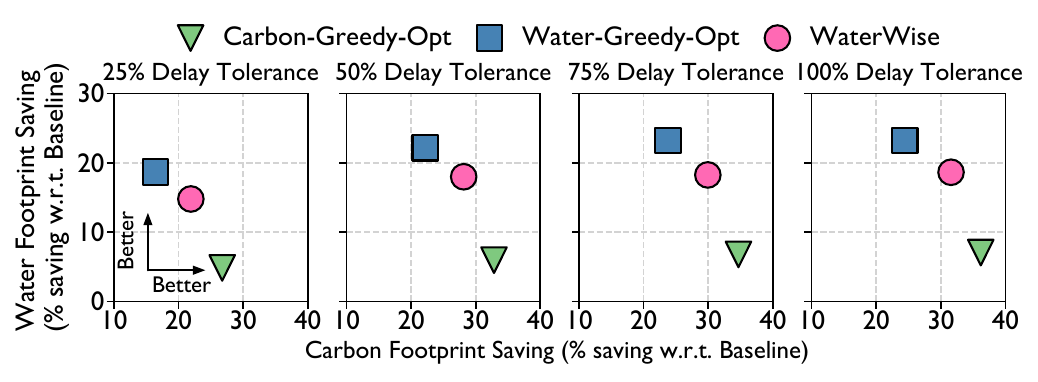}
    \vspace{-4mm}
    \caption{\sol{} provides significant carbon and water footprint savings compared to the baseline. Higher delay tolerance can further improve carbon and water footprint saving.}
    \label{fig:borg}
\end{figure}

\vspace{2mm}
\noindent\textbf{Figures of Merit.} The primary metrics in \sol{} are the carbon footprint and water footprint as defined in Sec.~\ref{sec: background}. They are represented as percentages of savings compared to the baseline (higher is better). We also report average execution time and the number of jobs with delay tolerance violations (as \% of all jobs), to better understand the performance, carbon, and water trade-offs, and \sol{}'s effectiveness. 

%% file: sections/evaluation.tex
\section{Evaluation}
\label{sec:evaluation}
\begin{figure}[t]
    \centering
    \includegraphics[scale=0.48]{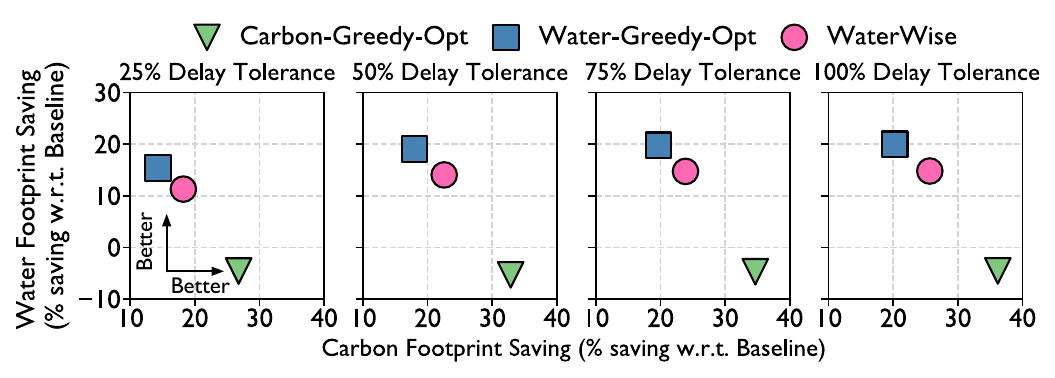}
    \vspace{-4mm}
    \caption{\todo{\sol{} consistently provides significant carbon and water footprint savings compared to the baseline, when evaluated with recent World Resources Institute data~\cite{reig2020guidance}.}}
    \label{fig:wri}
    \vspace{-4mm}
\end{figure}

\begin{figure}[t]
    \centering
    \includegraphics[scale=0.48]{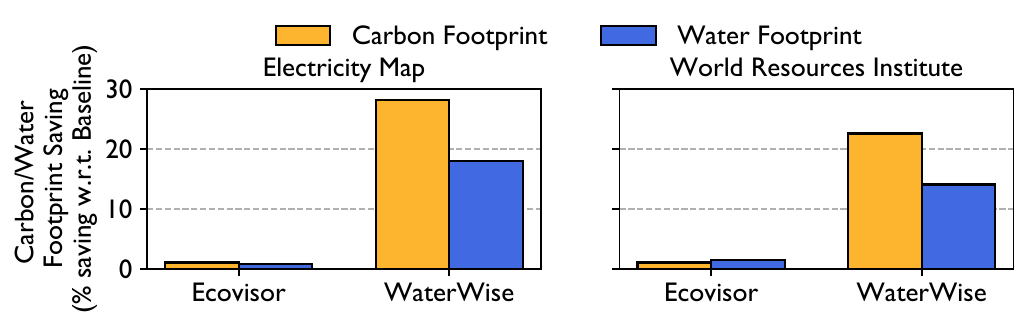}
    \vspace{-4mm}
    \caption{\todo{\sol{} yields higher carbon and water footprint savings compared to Ecovisor when evaluated with Electricity Map and World Resources Institute data. }}
    \label{fig:ecovisor}
    \vspace{-4mm}
\end{figure}

\sol{} provides significant carbon and water footprint savings, while operating under a delay tolerance constraint. Fig.~\ref{fig:borg} shows the effectiveness of \sol{} compared to Carbon-Greedy-Opt and Water-Greedy-Opt for different delay tolerances of service time when executing jobs following the Google Borg trace. 

We observe that \sol{} consistently balances between carbon footprint optimization and water footprint optimization and results in only 6.62\% carbon footprint increase with respect to Carbon-Greedy-Opt and 4.77\% water footprint increase with respect to Water-Greedy-Opt. As we increase the delay tolerance, jobs get more opportunities to be scheduled in remote data center locations for their execution, which increases their chances of further reducing carbon and water footprint. Thus, with an increase in delay tolerance, both \sol{}'s and Carbon-Greedy-Opt/Water-Greedy-Opt's performance improves. As the delay tolerance increases from 25\% to 100\%, the carbon footprint of \sol{} improves by 12.37\% and the water footprint improves by 4.52\%. Note that, \sol{} always performs better than the baseline solution by at least 21.91\% and 14.78\% in terms of carbon and water footprint, respectively, as the baseline solution always schedules jobs in their home data center regions in carbon-unaware and water-unaware fashion.

We acknowledge that embodied carbon footprint and water intensity calculation can have small inaccuracies or variations because the estimation relies on the accuracy of external data sources and methodology is not widely standardized yet (e.g., not all manufacturers report the embodied carbon consistently and not in the same way). Nevertheless, our sensitivity analysis shows that even with a 10\% variation in the estimation of embodied carbon and water intensity, \sol{} can still provide benefits. Specifically, \sol{} achieves 18\% and 26\% reduction in carbon and water footprint compared to the baseline solution, under a 10\% variation in embodied carbon and 50\% delay tolerance. Additionally, \sol{} reduces 28\% carbon footprint and 18\% water footprint within 50\% delay tolerance when subjected to a 10\% variation in water intensity.

\todo{In Fig.~\ref{fig:borg}, we used the energy mix breakdown data from the Electricity Maps~\cite{electricitymap} in all the evaluations. To evaluate the sensitivity of the water intensity data, we also compare using a different external dataset~\cite{reig2020guidance} from the World Resources Institute for calculating the water intensity in Fig.~\ref{fig:wri}. \sol{} yields over 18\% carbon footprint and 11\% water footprint savings compared to the baseline. }

\begin{figure}[t]
    \centering
    \includegraphics[scale=0.48]{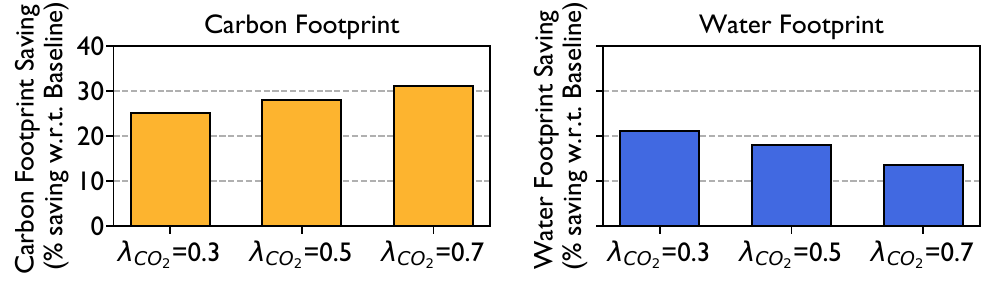}
    \vspace{-4mm}
    \caption{\todo{\sol{} is configurable and effective as the optimization weight factor for carbon and water footprint changes.}}
    \label{fig:lambda}
    \vspace{-4mm}
\end{figure}

\todo{In Fig.~\ref{fig:ecovisor}, we compare \sol{} with Ecovisor~\cite{souza2023ecovisor}. As discussed in Sec.\ref{sec:methodology}, Ecovisor focuses only on optimizing the carbon footprint, and not the water footprint — a key focus of \sol{}. The carbon scaler in Ecovisor adopts a flexible strategy that dynamically adjusts container resources, scaling up or down to maintain a consistent carbon footprint for long-running jobs, when the carbon intensity fluctuates. However, if the initial carbon intensity is high when the experiment begins, the target carbon footprint is always set high, resulting in an overall elevated carbon footprint for the job. Unlike \sol{}, Ecovisor primarily focuses on operational carbon during optimization, and the embodied carbon is not as thoroughly considered as \sol{}. Notably, embodied carbon is closely tied to the execution time of jobs, making it an essential component of comprehensive carbon-aware scheduling. Moreover, in the setting of Ecovisor, every job is executed in their home regions using Ecovisor, and Ecovisor does not explore cross-regional scheduling opportunities to reduce both carbon and water footprints at the same. Therefore, while Ecovisor is novel and effective for the purpose it was designed, it achieves modest carbon and water savings compared to the baseline scheme (but less than \sol{}). For example, using Electricity Map data, \sol{} reduces the carbon footprint by 27.6\% and the water footprint by 17.46\% compared to Ecovisor.}

\begin{figure}[t]
    \centering
    \includegraphics[scale=0.48]{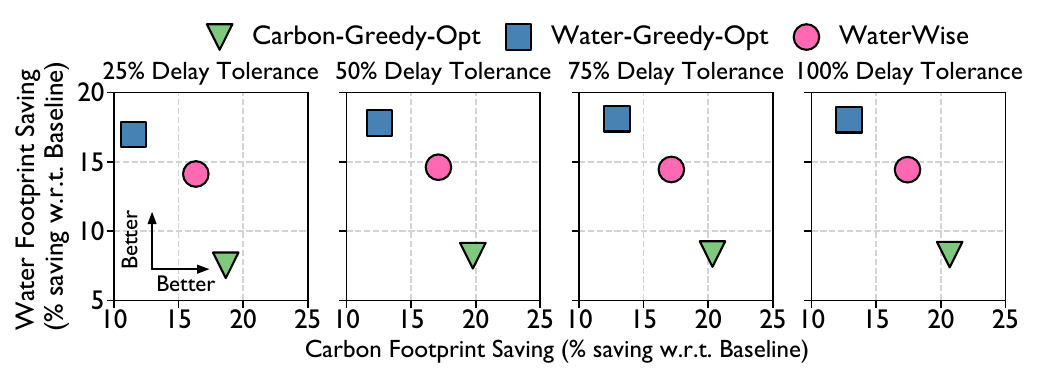}
    \vspace{-4mm}
    \caption{\sol{}'s effectiveness with the Alibaba trace.}
    \label{fig:alibaba}
    \vspace{-4mm}
\end{figure}

\todo{\sol{} remains effective even when the optimization weights for carbon and water footprints are reconfigured to be unequal. In Fig.~\ref{fig:lambda}, when carbon footprint minimization is given 30\% ($\lambda_{\text{CO}_2}=0.3$). \sol{} can still reduce 25.18\% carbon and 21.1\% water, within 50\% delay tolerance. Conversely, if carbon
footprint minimization is given 70\% ($\lambda_{\text{CO}_2}=0.7$). \sol{}
can still reduce 31.1\%	carbon and 13.6\% water compared to the baseline scheme, again within 50\% delay tolerance. }

\todo{\sol{} is designed to adapt to different request patterns and rates of uploading jobs. For example, even if the request rates in the Google Borg trace double compared to normal rate levels, \sol{} reduces carbon footprint by 21.7\% and water usage by 10.2\% compared to the baseline scheme. Additionally, } Fig.~\ref{fig:alibaba} shows similar trends in \sol{}'s performance when the jobs are invoked following \todo{the Alibaba Cloud VM  trace with different request patterns}. For example, on average, for a 25\% delay tolerance \sol{} performs within 3.43\% and 2.85\% of Carbon-Greedy-Opt and Water-Greedy-Opt in terms of carbon and water footprint, respectively. 

Recall that the major components of \sol{}'s design are the MILP solver and the slack manager, each of which plays a key role in \sol{}'s effectiveness. Based on the constraints of data center capacity in different regions and delay tolerances, \sol{}'s MILP solver schedules jobs to minimize carbon and water footprint, which is why \sol{}'s performance remains close to the optimal solutions. The slack manager considers the queue wait delay and transfer latency of jobs to different data center regions and accordingly forms solution constraints for the MILP solver. This is why, very few jobs violate the delay tolerance constraints as seen in Table~\ref{table:violations1}. In rare cases, where no feasible solution can be obtained by the MILP solver, the constraints are softened, which results in a few delay tolerance violations. As expected, the violations decrease with an increase in delay tolerance. Note that, these violations are not an artifact of \sol{}'s design -- in fact, even the optimal solutions experience these delay violations (Table~\ref{table:violations1}). They occur due to the limited capacity of data centers. 

\begin{table}
\centering
\caption{Average service time and delay tolerance violations.}
\vspace{-2mm}
\scalebox{0.65}{
\begin{tabular}{c|cccc|cccc}
\toprule
\hline
\multicolumn{1}{c|}{}& \multicolumn{4}{c|}{\makecell{\textbf{Service Time}\\(norm. to execution time)}}& \multicolumn{4}{c}{\textbf{\% Job Violation}}\\ 
\hline
\textbf{Delay Tolerance} & 25\% & 50\%  & 75\%   & 100\% & 25\% & 50\%  & 75\%   & 100\% \\ 
\hline
\multicolumn{1}{c|}{\cellcolor[HTML]{E6E6E6}Baseline}& \multicolumn{1}{c|}{\cellcolor[HTML]{E6E6E6}1$\times$} & \multicolumn{1}{c|}{\cellcolor[HTML]{E6E6E6}1$\times$} &  \multicolumn{1}{c|}{\cellcolor[HTML]{E6E6E6}1$\times$} &  \multicolumn{1}{c|}{\cellcolor[HTML]{E6E6E6}1$\times$} &  \multicolumn{1}{c|}{\cellcolor[HTML]{E6E6E6}0\%}  &  \multicolumn{1}{c|}{\cellcolor[HTML]{E6E6E6}0\%}  &  \multicolumn{1}{c|}{\cellcolor[HTML]{E6E6E6}0\%}  &  \multicolumn{1}{c}{\cellcolor[HTML]{E6E6E6}0\%} \\
\hline
 \multicolumn{1}{c|}{Carbon-Greedy-Opt} & \multicolumn{1}{c|}{1.06$\times$} & \multicolumn{1}{c|}{1.18$\times$} & \multicolumn{1}{c|}{1.28$\times$}& \multicolumn{1}{c|}{1.50$\times$} & \multicolumn{1}{c|}{1.1\%} & \multicolumn{1}{c|}{0.2\%} &\multicolumn{1}{c|}{0.3\%} & \multicolumn{1}{c}{0.5\%}\\
\hline
\multicolumn{1}{c|}{\cellcolor[HTML]{E6E6E6}Water-Greedy-Opt} & \multicolumn{1}{c|}{\cellcolor[HTML]{E6E6E6}1.07$\times$} & \multicolumn{1}{c|}{\cellcolor[HTML]{E6E6E6}1.17$\times$} & \multicolumn{1}{c|}{\cellcolor[HTML]{E6E6E6}1.23$\times$}& \multicolumn{1}{c|}{\cellcolor[HTML]{E6E6E6}1.34$\times$} & \multicolumn{1}{c|}{\cellcolor[HTML]{E6E6E6}1.7\%} & \multicolumn{1}{c|}{\cellcolor[HTML]{E6E6E6}0.1\%} & \multicolumn{1}{c|}{\cellcolor[HTML]{E6E6E6}0.2\%} & \multicolumn{1}{c}{\cellcolor[HTML]{E6E6E6}0.2\%}\\
\hline
\multicolumn{1}{c|}{\sol{}} & \multicolumn{1}{c|}{1.03$\times$} & \multicolumn{1}{c|}{1.09$\times$} & \multicolumn{1}{c|}{1.11$\times$}& \multicolumn{1}{c|}{1.13$\times$} & \multicolumn{1}{c|}{2.5\%} & \multicolumn{1}{c|}{0.03\%} & \multicolumn{1}{c|}{0.03\%} & \multicolumn{1}{c}{0.02\%}\\
\hline
\end{tabular}}
\vspace{-3mm}
\label{table:violations1}
\end{table}

Additionally, from Table~\ref{table:violations1} we also observe that the average service time is much lower than the delay tolerance. For example, for a 25\%, 50\%, 75\%, and 100\% service time delay tolerance compared to the execution time, the average delay of jobs is only 3\%, 9\%, 11\%, and 13\%, respectively.  This delay arises from the transfer latency of jobs to remote data centers and the queuing delay when data centers have no capacity for job execution. \sol{}'s slack manager designs constraints so that the MILP solver schedules jobs to minimize carbon and water footprint, without much increase in service time.  

\begin{figure}[t]
    \centering
    \includegraphics[scale=0.49]{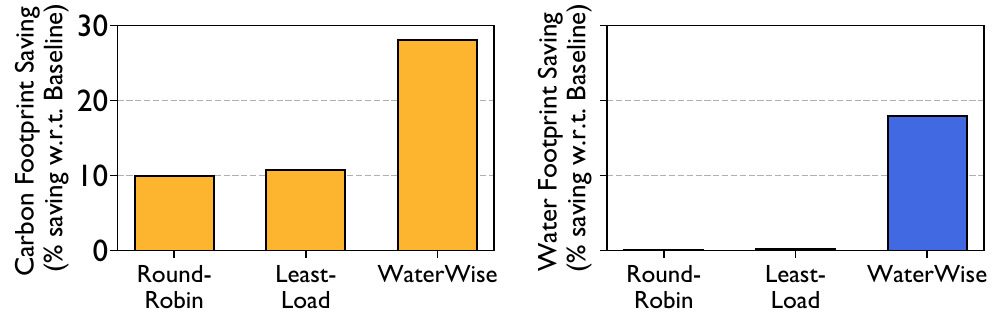}
    \vspace{-4mm}
    \caption{\sol{}'s comparison with other alternatives.}
    \vspace{-4mm}
    \label{fig:compare}
\end{figure}

Constraint-guided optimization of \sol{} to schedule jobs, helps it to obtain better solutions than other possible choices of scheduler design. In Fig.~\ref{fig:compare}, we compare the effectiveness of \sol{} with a scheduler that picks the data center region for job execution in a Round-Robin manner (Round-Robin), and one which schedules jobs in data center region with the least load (Least-Load). \sol{} performs scheduling based on the temporal and spatial variations of the factors affecting the carbon and water footprint (carbon intensity, EWIF, WUE, and WSF). This helps it to optimize carbon and water footprint by at least more than 19.5\%, and 17.8\%, respectively, compared to these carbon and water footprint-unaware scheduling solutions.

\begin{figure}[t]
    \centering
    \includegraphics[scale=0.5]{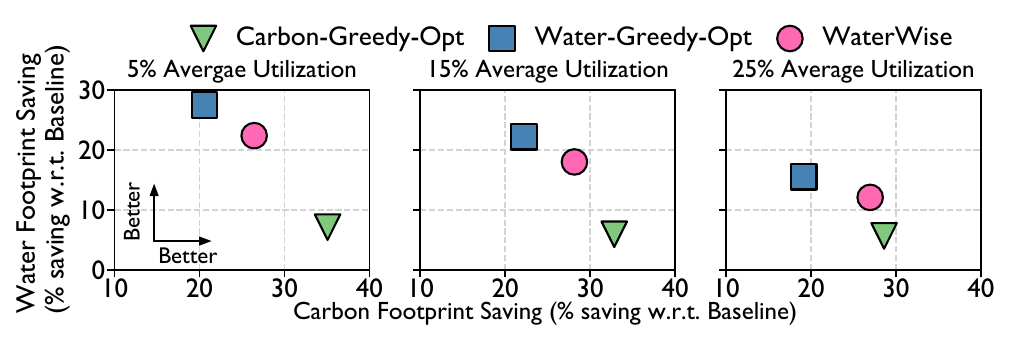}
    \vspace{-4mm}
    \caption{\sol{} across different utilization levels.}
    \label{fig:diff_util}
\end{figure}

All the previous results showed \sol{}'s performance with an average of 15\% server utilization in the different data center regions. In Fig.~\ref{fig:diff_util}, we show that \sol{} remains effective across different average utilization levels, which are obtained by changing the number of available servers in various data center regions.
For example, with a 5\% utilization, \sol{} performs with 13.31\% and 7.04\% of Carbon-Greedy-Opt and Water-Greedy-Opt in terms of carbon and water footprint, respectively. The capacity constraint changes based on the number of available servers with different data center utilization percentages. Based on this constraint, \sol{}'s MILP solver finds the optimal server for job execution.

\begin{figure}[t]
    \centering
    \includegraphics[scale=0.5]{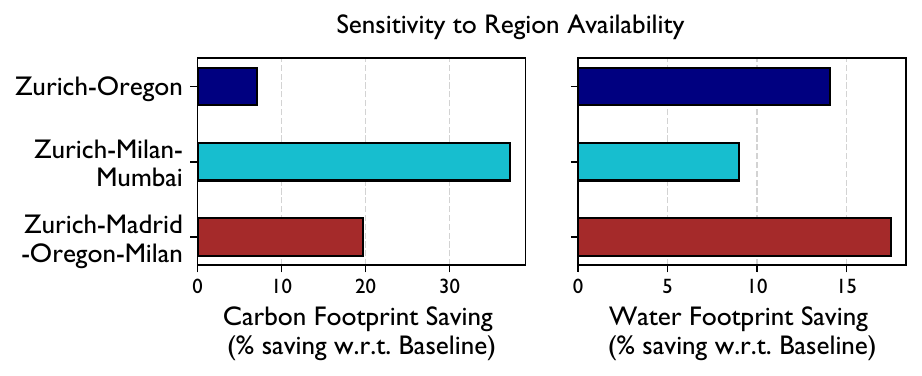}
    \vspace{-4mm}
    \caption{\todo{\sol{} is effective in saving carbon and water footprints under different resource availability.}}
    \label{fig:diff_regions}
\end{figure}

\todo{\sol{} can successfully adapt to changes in resource availability and remain effective. Fig.~\ref{fig:diff_regions} shows the results when we remove regions from the original evaluation setting. Different region availabilities can influence the carbon and water savings, due to the variance of carbon and water intensities. For example, the availability of Mumbai enables significant carbon savings, this is because \sol{} can explore the variances of carbon intensities across these 3 regions (Zurich, Milan, Mumbai). A job submitted from Mumbai (high carbon intensity region) can save carbon footprint, if this job is scheduled to Zurich (low carbon intensity region).}

\begin{figure}[t]
    \centering
    \includegraphics[scale=0.5]{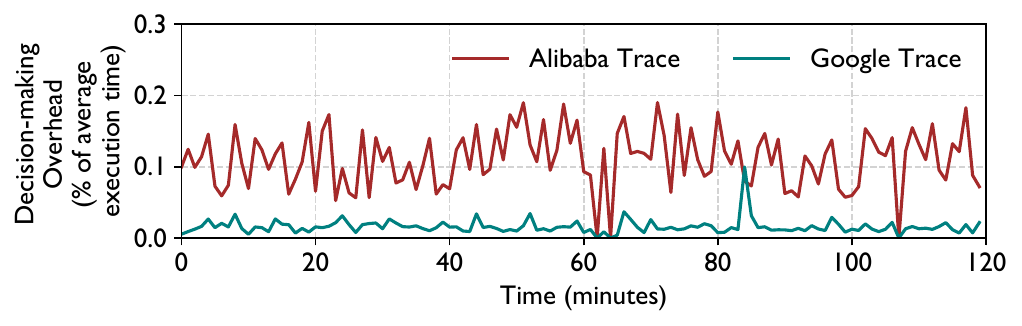}
    \vspace{-4mm}
    \caption{\sol{} demonstrates low overhead.}
        \vspace{-4mm}
    \label{fig:overhead}
\end{figure}

Finally, we demonstrate that the decision-making overhead of \sol{} is practical and minimal. Fig.~\ref{fig:overhead} shows that \sol{} decision-making overhead is negligible -- less than 0.2\% of the average execution time of the jobs. This is because \sol{} uses MILP solver which minimizes overhead by the use of solutions such as branch-and-cut to dynamically add valid inequalities to tighten the relaxation of the problem, and through efficient data structures like sparse matrices to reduce memory and computational costs. Additionally, it distributes the computation over multiple cores, accelerating convergence. Since the Alibaba trace has an 8.5$\times$ more job invocation rate than the Google Borg trace, we observe a higher decision-making overhead when invoking jobs following the Alibaba trace. The temporal variations in decision-making overhead are due to the variations in job invocation rate following Google Borg or Alibaba trace.

\begin{table}
\centering
\caption{\todo{Communication overhead (the home region is Oregon).}}
\vspace{-2mm}
\scalebox{0.65}{
\begin{tabular}{c|c|c}
\toprule
\hline
\multicolumn{1}{c|}{\textbf{Regions}}& \multicolumn{1}{c|}{\makecell{\textbf{Avg Carbon Overhead}\\(\% Execution Carbon)}}& \multicolumn{1}{c}{\makecell{\textbf{Avg Water Overhead}\\(\% Execution Water)}}\\ 
\hline
\multicolumn{1}{c|}{\cellcolor[HTML]{E6E6E6}Zurich}& \multicolumn{1}{c|}{\cellcolor[HTML]{E6E6E6}0.08} & \multicolumn{1}{c}{\cellcolor[HTML]{E6E6E6}0.09}\\
\hline
 \multicolumn{1}{c|}{Madrid} & \multicolumn{1}{c|}{0.08} & \multicolumn{1}{c}{0.09} \\
\hline
\multicolumn{1}{c|}{\cellcolor[HTML]{E6E6E6}Milan}  & \multicolumn{1}{c|}{\cellcolor[HTML]{E6E6E6}0.13} & \multicolumn{1}{c}{\cellcolor[HTML]{E6E6E6}0.09}\\
\hline
\multicolumn{1}{c|}{Mumbai} & \multicolumn{1}{c|}{0.17} & \multicolumn{1}{c}{0.13} \\
\hline
\end{tabular}}
\vspace{-3mm}
\label{table:communication_cost}
\end{table}

\todo{The delay tolerance component in \sol{} accounts for the communication costs of sending and executing the job to a remote location (including the application package transfer). In Table~\ref{table:communication_cost}, we present an example result where the home location is Oregon and the results are needed locally after execution at a remote location, it adds only small latency, carbon, and water footprint overhead, expressed as percentages of the execution time, the carbon footprint during execution, and the water footprint during execution, respectively. Transfer latency takes the majority part in the communication cost. }

%% file: sections/discussion.tex
\section{Discussion}
\label{sec:discussion}
\vspace{2mm}
\noindent\textbf{Generalization to Different Data Centers.} \todo{\sol{}’s carbon and water optimizations are inherently designed to capture the key sustainability factors that can vary across data centers, such as PUE, carbon emissions, and onsite \& offsite water consumption (EWIF, WUE). By incorporating these parameters, \sol{} allows for comprehensive analysis and informed decision-making to reduce environmental impact. Many individual data center parameters such as PUE, CI, EWIF, WUE, and WSF are configurable in the \sol{} framework for the community to gain insights into water and carbon trade-offs. While \sol{}'s evaluation focuses on a set of representative parameters, these configurable parameters allow us to evaluate the impact across data centers.}

\vspace{2mm}
\noindent\textbf{Performance Considerations.} \todo{\sol{}’s primary focus is on co-optimizing carbon and water, instead of pushing the state-of-the-art for scheduling for performance for executing jobs among different regions. \sol{} is complementary to performance-only focused scheduling and does not inherently prohibit integration with such techniques. For any technique that focuses on performance only,\sol{} can be configured to treat performance as a third objective with a configurable weight factor instead of our current delay-tolerance feature, which accounts for allowable scheduling slack based on environmental goals, could serve as a foundation for incorporating performance-driven criteria. This approach can broaden the scope of scheduling systems -- the integration of sustainability with traditional computational goals.}

\vspace{2mm}
\noindent\textbf{Cost Considerations.} \todo{The primary goal of \sol{} is to demonstrate that carbon and water are competing sustainability metrics. Similar to how performance considerations, incorporating financial costs, such as energy consumption or VM usage, into the optimization objective could dilute the environmental focus and add unnecessary complexity to the model, especially given the variability of cost structures across providers and regions. However, financial costs can be integrated into the optimization goals in future extensions to further enhance the \sol{}. }

%% file: sections/related_work.tex
\section{Related Work}
\label{sec:related}
Optimizing for performance metrics like latency and throughput, as well as designing energy-efficient hardware and software stacks, have long been central to data center research ~\cite{bambagini2016energy, petrucci2015octopus, goiri2011greenslot,gao2020smartly,agrawal2014energy}.
However, optimizing solely for energy efficiency is insufficient to minimize the environmental impact of data centers. 
Sustainability metrics, such as carbon and water footprints, have largely been overlooked in data center research until recently when there has been a growing interest in addressing this gap~\cite{gupta2021chasing, li2023making, monserrate2022cloud, li2023toward, 2023wangpeelingback}. Optimizing for energy is not equivalent to optimizing for carbon footprint as the latter also depends on carbon intensity and the embodied carbon of a device. Researchers have proposed strategies to scale resources allocated to applications to match the carbon cost of the energy supply, exploit temporal variation in carbon intensity~\cite{hanafy2023carbonscaler,wiesner2021waitawhile,souza2023ecovisor}, or incorporate geographical load balancing strategies~\cite{liu2015greening,2023majicarbonmulticloud} or both~\cite{2024limitationssukprasert}.  
Focusing solely on the carbon-performance tradeoff can increase application running costs. To address this, GAIA, a carbon-aware scheduler introduced in \cite{hanafy2024gaia}, co-optimizes for performance, carbon emissions, and cost in cloud-based batch processing systems.

Efficient server design using carbon-efficient server components is another opportunity that previous research has explored~\cite{frachtenberg2011highefficiency, mirhosseini2019killermicro, wang2024designing, elgamal2023carbonefficientdesign, acun2023carbon}. Strategical placement of data centers to minimize one or more environmental footprints is another approach that has been proposed~\cite{siddik2021datacenterfootprint}.

In~\cite{li2023making}, the authors provide a methodology to estimate the water footprint of AI models and show that spatial and temporal diversity of WUE can be leveraged to reduce the water footprint of AI model training. Water-focused or water-aware research is beginning to gain attention~\cite{islam2018wace,zuccon2023beyond,li2023environmentally,egbemhenghe2023revolutionizing}.  For example, the water consumption of information retrieval systems is analyzed in~\cite{zuccon2023beyond}.
In~\cite{li2023environmentally}, the authors analyze the carbon and water footprints of AI model inference. However, no prior work systematically exploits the co-optimization of carbon and water footprint -- esp. considering the water scarcity factor. As we show in this work, simply saving water consumption is not sufficient -- it is equally critical to understand and consider the regional preciousness of water and understand the water generation trade-offs for carbon-friendly energy sources for electricity generation. While the community has access to carbon-aware resource management frameworks, it does not have access to open-source frameworks and open-source water data to develop new carbon-aware and water-aware resource management strategies -- this work will bridge that gap.

%% file: sections/conclusion.tex
\section{Conclusion}
\label{sec:conclusion}

This paper introduced \sol{}, a framework designed to co-optimize the carbon and water footprint of parallel jobs in geographically distributed data centers. This paper comprehensively analyzes and models the carbon and water footprint to reveal new insights and design a novel scheduler. \sol{} provides approx. 21\% carbon-footprint savings and approx. 14\% water-footprint savings compared to the carbon- and water-unaware scheduler.

\vspace{1mm}
\noindent\textbf{Acknowledgment}. We thank Yuanchao Xu (our shepherd) and the reviewers for their constructive feedback. This
work is supported by NSF Awards 1910601, 2124897.

%% file: sections/artifact.tex
\appendix
\section{Artifact Appendix}

\subsection{Abstract}
\sol{} is an innovative job scheduling framework designed to co-optimize the carbon and water footprints of geographically distributed data centers. It leverages a Mixed Integer Linear Programming (MILP) approach to dynamically adjust scheduling based on temporal and spatial variations in carbon intensity and water intensity. \sol{} balances the optimization goals with low decision-making overhead and delay latency by incorporating delay tolerance and slack management. This open-source artifact is available at:

\begin{center}
    \url{https://zenodo.org/records/14219862}
\end{center}

The artifact includes the following:
\begin{itemize}
    \item Data for sustainable metrics, including WUE, EWIF, WSF, and carbon intensity.
    \item The profiling data for energy consumption, and execution time of parallel applications.
    \item Scripts and data for generating motivation plots.
    \item Scripts to setting up and evaluating the \sol{}.
    \item Scripts for implementing and comparing different optimizers within \sol{}.
\end{itemize}
\subsection{Artifact check-list (meta-information)}
{\small
\begin{itemize}
  \item {\bf Algorithm: }A MILP-based algorithm for scheduling parallel workloads.
  \item {\bf Program: }Optimizers designed to reduce carbon and water footprints.
  \item {\bf Data set: }Carbon intensity and EWIF are sourced from Electricity Map, while WUE is calculated based on wet bulb temperature data from Meteologix. More details are provided in the artifact.
  \item {\bf Run-time environment: }Python 3.8 with required Python packages.
  \item {\bf Hardware: }No specific hardware is required for simulation. Profiling data is collected from AWS EC2 m5.metal bare-metal machines.
  \item {\bf Execution: }Follow the instructions in README.md.
  \item {\bf Metrics: }Includes execution time, energy consumption, carbon footprint, water footprint for a single job, and transfer latency between regions.
  \item {\bf Output: }Logs are generated, including carbon and water footprint, execution time, and overhead. Refer to README.md for more details.
  \item {\bf Experiments: }(1). Experiments using Google Borg trace. (2). Experiments using Alibaba VM trace. (3). Sensitivity and robustness analysis.
  \item {\bf How much disk space required (approximately)?: }More than 10 GB.
  \item {\bf How much time is needed to prepare workflow (approximately)?: }5 mins.
  \item {\bf How much time is needed to complete experiments (approximately)?: }More than 1 hour.
  \item {\bf Publicly available?: }Yes
  \item {\bf Archived (DOI)?: }https://zenodo.org/records/14219862
\end{itemize}
}

\subsection{Description}
This artifact provides the framework of \sol{}, designed to execute parallel jobs on cloud servers. \sol{} simulates job execution using industry-standard VM cloud traces. The inputs include carbon intensity, water intensity, and job data. After configuring the parameters of \sol{}, the framework performs a 10-day simulation based on the VM trace (230,000 jobs from the Google Borg trace). \sol{}'s Decision Controller leverages a MILP algorithm to determine which jobs are sent to which regions, generating detailed log files for reference.

\subsection{Installation}
To set up the \sol{} artifact, refer to the README.md file included in the artifact. The README.md provides detailed installation instructions. Additionly, several Python libraries need to be installed to run the artifact.

\subsection{Expected results}

Main figures are expected to be reproduced from the data in the artifact.

%% file: main.bbl

\begin{thebibliography}{61}


\ifx \showCODEN    \undefined \def \showCODEN     #1{\unskip}     \fi
\ifx \showDOI      \undefined \def \showDOI       #1{#1}\fi
\ifx \showISBNx    \undefined \def \showISBNx     #1{\unskip}     \fi
\ifx \showISBNxiii \undefined \def \showISBNxiii  #1{\unskip}     \fi
\ifx \showISSN     \undefined \def \showISSN      #1{\unskip}     \fi
\ifx \showLCCN     \undefined \def \showLCCN      #1{\unskip}     \fi
\ifx \shownote     \undefined \def \shownote      #1{#1}          \fi
\ifx \showarticletitle \undefined \def \showarticletitle #1{#1}   \fi
\ifx \showURL      \undefined \def \showURL       {\relax}        \fi
\providecommand\bibfield[2]{#2}
\providecommand\bibinfo[2]{#2}
\providecommand\natexlab[1]{#1}
\providecommand\showeprint[2][]{arXiv:#2}

\bibitem[our(2024)]%
        {ourworldindataWaterStress}
 \bibinfo{year}{2024}\natexlab{}.
\newblock \bibinfo{title}{{W}ater {U}se and {S}tress --- ourworldindata.org}.
\newblock \bibinfo{howpublished}{\url{https://ourworldindata.org/water-use-stress}}.
\newblock


\bibitem[Acun et~al\mbox{.}(2023)]%
        {acun2023carbon}
\bibfield{author}{\bibinfo{person}{Bilge Acun}, \bibinfo{person}{Benjamin Lee}, \bibinfo{person}{Fiodar Kazhamiaka}, \bibinfo{person}{Aditya Sundarrajan}, \bibinfo{person}{Kiwan Maeng}, \bibinfo{person}{Manoj Chakkaravarthy}, \bibinfo{person}{David Brooks}, {and} \bibinfo{person}{Carole-Jean Wu}.} \bibinfo{year}{2023}\natexlab{}.
\newblock \showarticletitle{Carbon dependencies in datacenter design and management}.
\newblock \bibinfo{journal}{\emph{ACM SIGENERGY Energy Informatics Review}} \bibinfo{volume}{3}, \bibinfo{number}{3} (\bibinfo{year}{2023}), \bibinfo{pages}{21--26}.
\newblock


\bibitem[Agrawal and Rao(2014)]%
        {agrawal2014energy}
\bibfield{author}{\bibinfo{person}{Pragati Agrawal} {and} \bibinfo{person}{Shrisha Rao}.} \bibinfo{year}{2014}\natexlab{}.
\newblock \showarticletitle{Energy-aware scheduling of distributed systems}.
\newblock \bibinfo{journal}{\emph{IEEE Transactions on Automation Science and Engineering}} \bibinfo{volume}{11}, \bibinfo{number}{4} (\bibinfo{year}{2014}), \bibinfo{pages}{1163--1175}.
\newblock


\bibitem[Ambati et~al\mbox{.}(2021)]%
        {ambati2021good}
\bibfield{author}{\bibinfo{person}{Pradeep Ambati}, \bibinfo{person}{Noman Bashir}, \bibinfo{person}{David Irwin}, {and} \bibinfo{person}{Prashant Shenoy}.} \bibinfo{year}{2021}\natexlab{}.
\newblock \showarticletitle{Good things come to those who wait: Optimizing job waiting in the cloud}. In \bibinfo{booktitle}{\emph{Proceedings of the ACM Symposium on Cloud Computing}}. \bibinfo{pages}{229--242}.
\newblock


\bibitem[Anderson et~al\mbox{.}(2023)]%
        {anderson2023treehouse}
\bibfield{author}{\bibinfo{person}{Thomas Anderson}, \bibinfo{person}{Adam Belay}, \bibinfo{person}{Mosharaf Chowdhury}, \bibinfo{person}{Asaf Cidon}, {and} \bibinfo{person}{Irene Zhang}.} \bibinfo{year}{2023}\natexlab{}.
\newblock \showarticletitle{Treehouse: A case for carbon-aware datacenter software}.
\newblock \bibinfo{journal}{\emph{ACM SIGENERGY Energy Informatics Review}} \bibinfo{volume}{3}, \bibinfo{number}{3} (\bibinfo{year}{2023}), \bibinfo{pages}{64--70}.
\newblock


\bibitem[Auger et~al\mbox{.}(2021)]%
        {auger2021open}
\bibfield{author}{\bibinfo{person}{Cl{\'e}ment Auger}, \bibinfo{person}{Benoit Hilloulin}, \bibinfo{person}{Benjamin Boisserie}, \bibinfo{person}{Ma{\"e}l Thomas}, \bibinfo{person}{Quentin Guignard}, {and} \bibinfo{person}{Emmanuel Rozi{\`e}re}.} \bibinfo{year}{2021}\natexlab{}.
\newblock \showarticletitle{Open-source carbon footprint estimator: Development and university declination}.
\newblock \bibinfo{journal}{\emph{Sustainability}} \bibinfo{volume}{13}, \bibinfo{number}{8} (\bibinfo{year}{2021}), \bibinfo{pages}{4315}.
\newblock


\bibitem[Bambagini et~al\mbox{.}(2016)]%
        {bambagini2016energy}
\bibfield{author}{\bibinfo{person}{Mario Bambagini}, \bibinfo{person}{Mauro Marinoni}, \bibinfo{person}{Hakan Aydin}, {and} \bibinfo{person}{Giorgio Buttazzo}.} \bibinfo{year}{2016}\natexlab{}.
\newblock \showarticletitle{Energy-aware scheduling for real-time systems: A survey}.
\newblock \bibinfo{journal}{\emph{ACM Transactions on Embedded Computing Systems (TECS)}} \bibinfo{volume}{15}, \bibinfo{number}{1} (\bibinfo{year}{2016}), \bibinfo{pages}{1--34}.
\newblock


\bibitem[Boyd and Vandenberghe(2004)]%
        {boyd2004convex}
\bibfield{author}{\bibinfo{person}{Stephen Boyd} {and} \bibinfo{person}{Lieven Vandenberghe}.} \bibinfo{year}{2004}\natexlab{}.
\newblock \bibinfo{booktitle}{\emph{Convex optimization}}.
\newblock \bibinfo{publisher}{Cambridge university press}.
\newblock


\bibitem[Bruckner et~al\mbox{.}(2014)]%
        {bruckner2014technology}
\bibfield{author}{\bibinfo{person}{Thomas Bruckner}, \bibinfo{person}{L Fulton}, \bibinfo{person}{E Hertwich}, \bibinfo{person}{A McKinnon}, \bibinfo{person}{D Perczyk}, \bibinfo{person}{J Roy}, \bibinfo{person}{R Schaeffer}, \bibinfo{person}{S Schl{\"o}mer}, \bibinfo{person}{R Sims}, \bibinfo{person}{P Smith}, {and} \bibinfo{person}{R Wiser}.} \bibinfo{year}{2014}\natexlab{}.
\newblock \showarticletitle{Technology-specific cost and performance parameters [annex III]}.
\newblock In \bibinfo{booktitle}{\emph{Climate change 2014: mitigation of climate change}}. \bibinfo{publisher}{Cambridge University Press}, \bibinfo{pages}{1329--1356}.
\newblock


\bibitem[Chen et~al\mbox{.}(2019)]%
        {chen2019parties}
\bibfield{author}{\bibinfo{person}{Shuang Chen}, \bibinfo{person}{Christina Delimitrou}, {and} \bibinfo{person}{Jos{\'e}~F Mart{\'\i}nez}.} \bibinfo{year}{2019}\natexlab{}.
\newblock \showarticletitle{Parties: Qos-aware resource partitioning for multiple interactive services}. In \bibinfo{booktitle}{\emph{Proceedings of the Twenty-Fourth International Conference on Architectural Support for Programming Languages and Operating Systems}}. \bibinfo{pages}{107--120}.
\newblock


\bibitem[Cheng et~al\mbox{.}(2018)]%
        {cheng2018analyzing}
\bibfield{author}{\bibinfo{person}{Yue Cheng}, \bibinfo{person}{Ali Anwar}, {and} \bibinfo{person}{Xuejing Duan}.} \bibinfo{year}{2018}\natexlab{}.
\newblock \showarticletitle{Analyzing alibaba’s co-located datacenter workloads}. In \bibinfo{booktitle}{\emph{2018 IEEE International Conference on Big Data (Big Data)}}. IEEE, \bibinfo{pages}{292--297}.
\newblock


\bibitem[Davy(2021)]%
        {davy2021building}
\bibfield{author}{\bibinfo{person}{Benjamin Davy}.} \bibinfo{year}{2021}\natexlab{}.
\newblock \showarticletitle{Building an AWS EC2 carbon emissions dataset}.
\newblock \bibinfo{journal}{\emph{Medium, September}} (\bibinfo{year}{2021}).
\newblock


\bibitem[Davy(2024)]%
        {Davy_2024}
\bibfield{author}{\bibinfo{person}{Benjamin Davy}.} \bibinfo{year}{2024}\natexlab{}.
\newblock \bibinfo{title}{Building an AWS EC2 carbon emissions dataset}.
\newblock
\newblock
\urldef\tempurl%
\url{https://medium.com/teads-engineering/building-an-aws-ec2-carbon-emissions-dataset-3f0fd76c98ac}
\showURL{%
\tempurl}


\bibitem[Ebrahimi et~al\mbox{.}(2014)]%
        {ebrahimi2014review}
\bibfield{author}{\bibinfo{person}{Khosrow Ebrahimi}, \bibinfo{person}{Gerard~F Jones}, {and} \bibinfo{person}{Amy~S Fleischer}.} \bibinfo{year}{2014}\natexlab{}.
\newblock \showarticletitle{A review of data center cooling technology, operating conditions and the corresponding low-grade waste heat recovery opportunities}.
\newblock \bibinfo{journal}{\emph{Renewable and sustainable energy reviews}}  \bibinfo{volume}{31} (\bibinfo{year}{2014}), \bibinfo{pages}{622--638}.
\newblock


\bibitem[Eeckhout(2022)]%
        {eeckhout2022first}
\bibfield{author}{\bibinfo{person}{Lieven Eeckhout}.} \bibinfo{year}{2022}\natexlab{}.
\newblock \showarticletitle{A first-order model to assess computer architecture sustainability}.
\newblock \bibinfo{journal}{\emph{IEEE Computer Architecture Letters}} \bibinfo{volume}{21}, \bibinfo{number}{2} (\bibinfo{year}{2022}), \bibinfo{pages}{137--140}.
\newblock


\bibitem[Egbemhenghe et~al\mbox{.}(2023)]%
        {egbemhenghe2023revolutionizing}
\bibfield{author}{\bibinfo{person}{Abel~U Egbemhenghe}, \bibinfo{person}{Toluwalase Ojeyemi}, \bibinfo{person}{Kingsley~O Iwuozor}, \bibinfo{person}{Ebuka~Chizitere Emenike}, \bibinfo{person}{Tolu~I Ogunsanya}, \bibinfo{person}{Stella~Ukamaka Anidiobi}, {and} \bibinfo{person}{Adewale~George Adeniyi}.} \bibinfo{year}{2023}\natexlab{}.
\newblock \showarticletitle{Revolutionizing water treatment, conservation, and management: Harnessing the power of AI-driven ChatGPT solutions}.
\newblock \bibinfo{journal}{\emph{Environmental Challenges}}  \bibinfo{volume}{13} (\bibinfo{year}{2023}), \bibinfo{pages}{100782}.
\newblock


\bibitem[Elgamal et~al\mbox{.}(2023)]%
        {elgamal2023carbonefficientdesign}
\bibfield{author}{\bibinfo{person}{Mariam Elgamal}, \bibinfo{person}{Doug Carmean}, \bibinfo{person}{Elnaz Ansari}, \bibinfo{person}{Okay Zed}, \bibinfo{person}{Ramesh Peri}, \bibinfo{person}{Srilatha Manne}, \bibinfo{person}{Udit Gupta}, \bibinfo{person}{Gu-Yeon Wei}, \bibinfo{person}{David Brooks}, \bibinfo{person}{Gage Hills}, {and} \bibinfo{person}{Carole-Jean Wu}.} \bibinfo{year}{2023}\natexlab{}.
\newblock \showarticletitle{Carbon-Efficient Design Optimization for Computing Systems}. In \bibinfo{booktitle}{\emph{Proceedings of the 2nd Workshop on Sustainable Computer Systems}} (Boston, MA, USA) \emph{(\bibinfo{series}{HotCarbon '23})}. \bibinfo{publisher}{Association for Computing Machinery}, \bibinfo{address}{New York, NY, USA}, Article \bibinfo{articleno}{16}, \bibinfo{numpages}{7}~pages.
\newblock
\showISBNx{9798400702426}
\urldef\tempurl%
\url{https://doi.org/10.1145/3604930.3605712}
\showDOI{\tempurl}


\bibitem[Ferdman et~al\mbox{.}(2012)]%
        {ferdman2012clearing}
\bibfield{author}{\bibinfo{person}{Michael Ferdman}, \bibinfo{person}{Almutaz Adileh}, \bibinfo{person}{Onur Kocberber}, \bibinfo{person}{Stavros Volos}, \bibinfo{person}{Mohammad Alisafaee}, \bibinfo{person}{Djordje Jevdjic}, \bibinfo{person}{Cansu Kaynak}, \bibinfo{person}{Adrian~Daniel Popescu}, \bibinfo{person}{Anastasia Ailamaki}, {and} \bibinfo{person}{Babak Falsafi}.} \bibinfo{year}{2012}\natexlab{}.
\newblock \showarticletitle{Clearing the clouds: a study of emerging scale-out workloads on modern hardware}.
\newblock \bibinfo{journal}{\emph{Acm sigplan notices}} \bibinfo{volume}{47}, \bibinfo{number}{4} (\bibinfo{year}{2012}), \bibinfo{pages}{37--48}.
\newblock


\bibitem[Floudas and Lin(2005)]%
        {floudas2005mixed}
\bibfield{author}{\bibinfo{person}{Christodoulos~A Floudas} {and} \bibinfo{person}{Xiaoxia Lin}.} \bibinfo{year}{2005}\natexlab{}.
\newblock \showarticletitle{Mixed integer linear programming in process scheduling: Modeling, algorithms, and applications}.
\newblock \bibinfo{journal}{\emph{Annals of Operations Research}}  \bibinfo{volume}{139} (\bibinfo{year}{2005}), \bibinfo{pages}{131--162}.
\newblock


\bibitem[Frachtenberg et~al\mbox{.}(2011)]%
        {frachtenberg2011highefficiency}
\bibfield{author}{\bibinfo{person}{Eitan Frachtenberg}, \bibinfo{person}{Ali Heydari}, \bibinfo{person}{Harry Li}, \bibinfo{person}{Amir Michael}, \bibinfo{person}{Jacob Na}, \bibinfo{person}{Avery Nisbet}, {and} \bibinfo{person}{Pierluigi Sarti}.} \bibinfo{year}{2011}\natexlab{}.
\newblock \showarticletitle{High-efficiency server design}. In \bibinfo{booktitle}{\emph{Proceedings of 2011 International Conference for High Performance Computing, Networking, Storage and Analysis}} (Seattle, Washington) \emph{(\bibinfo{series}{SC '11})}. \bibinfo{publisher}{Association for Computing Machinery}, \bibinfo{address}{New York, NY, USA}, Article \bibinfo{articleno}{27}, \bibinfo{numpages}{27}~pages.
\newblock
\showISBNx{9781450307710}
\urldef\tempurl%
\url{https://doi.org/10.1145/2063384.2063420}
\showDOI{\tempurl}


\bibitem[Gao et~al\mbox{.}(2020)]%
        {gao2020smartly}
\bibfield{author}{\bibinfo{person}{Jiechao Gao}, \bibinfo{person}{Haoyu Wang}, {and} \bibinfo{person}{Haiying Shen}.} \bibinfo{year}{2020}\natexlab{}.
\newblock \showarticletitle{Smartly handling renewable energy instability in supporting a cloud datacenter}. In \bibinfo{booktitle}{\emph{2020 IEEE international parallel and distributed processing symposium (IPDPS)}}. IEEE, \bibinfo{pages}{769--778}.
\newblock


\bibitem[Goiri et~al\mbox{.}(2011)]%
        {goiri2011greenslot}
\bibfield{author}{\bibinfo{person}{{\'I}{\~n}igo Goiri}, \bibinfo{person}{Kien Le}, \bibinfo{person}{Md~E Haque}, \bibinfo{person}{Ryan Beauchea}, \bibinfo{person}{Thu~D Nguyen}, \bibinfo{person}{Jordi Guitart}, \bibinfo{person}{Jordi Torres}, {and} \bibinfo{person}{Ricardo Bianchini}.} \bibinfo{year}{2011}\natexlab{}.
\newblock \showarticletitle{Greenslot: scheduling energy consumption in green datacenters}. In \bibinfo{booktitle}{\emph{Proceedings of 2011 International Conference for High Performance Computing, Networking, Storage and Analysis}}. \bibinfo{pages}{1--11}.
\newblock


\bibitem[Google(2023)]%
        {googlereport}
\bibfield{author}{\bibinfo{person}{Google}.} \bibinfo{year}{2023}\natexlab{}.
\newblock \bibinfo{booktitle}{\emph{Google 2023 Environmental Repor}}.
\newblock
\urldef\tempurl%
\url{https://www.gstatic.com/gumdrop/sustainability/google-2023-environmental-report.pdf}
\showURL{%
\tempurl}


\bibitem[Gupta et~al\mbox{.}(2022)]%
        {gupta2022act}
\bibfield{author}{\bibinfo{person}{Udit Gupta}, \bibinfo{person}{Mariam Elgamal}, \bibinfo{person}{Gage Hills}, \bibinfo{person}{Gu-Yeon Wei}, \bibinfo{person}{Hsien-Hsin~S Lee}, \bibinfo{person}{David Brooks}, {and} \bibinfo{person}{Carole-Jean Wu}.} \bibinfo{year}{2022}\natexlab{}.
\newblock \showarticletitle{ACT: Designing sustainable computer systems with an architectural carbon modeling tool}. In \bibinfo{booktitle}{\emph{Proceedings of the 49th Annual International Symposium on Computer Architecture}}. \bibinfo{pages}{784--799}.
\newblock


\bibitem[Gupta et~al\mbox{.}(2021)]%
        {gupta2021chasing}
\bibfield{author}{\bibinfo{person}{Udit Gupta}, \bibinfo{person}{Young~Geun Kim}, \bibinfo{person}{Sylvia Lee}, \bibinfo{person}{Jordan Tse}, \bibinfo{person}{Hsien-Hsin~S Lee}, \bibinfo{person}{Gu-Yeon Wei}, \bibinfo{person}{David Brooks}, {and} \bibinfo{person}{Carole-Jean Wu}.} \bibinfo{year}{2021}\natexlab{}.
\newblock \showarticletitle{Chasing carbon: The elusive environmental footprint of computing}. In \bibinfo{booktitle}{\emph{2021 IEEE International Symposium on High-Performance Computer Architecture (HPCA)}}. IEEE, \bibinfo{pages}{854--867}.
\newblock


\bibitem[Hanafy et~al\mbox{.}(2023)]%
        {hanafy2023carbonscaler}
\bibfield{author}{\bibinfo{person}{Walid~A Hanafy}, \bibinfo{person}{Qianlin Liang}, \bibinfo{person}{Noman Bashir}, \bibinfo{person}{David Irwin}, {and} \bibinfo{person}{Prashant Shenoy}.} \bibinfo{year}{2023}\natexlab{}.
\newblock \showarticletitle{CarbonScaler: Leveraging Cloud Workload Elasticity for Optimizing Carbon-Efficiency}.
\newblock \bibinfo{journal}{\emph{arXiv preprint arXiv:2302.08681}} (\bibinfo{year}{2023}).
\newblock


\bibitem[Hanafy et~al\mbox{.}(2024)]%
        {hanafy2024gaia}
\bibfield{author}{\bibinfo{person}{Walid~A. Hanafy}, \bibinfo{person}{Qianlin Liang}, \bibinfo{person}{Noman Bashir}, \bibinfo{person}{Abel Souza}, \bibinfo{person}{David Irwin}, {and} \bibinfo{person}{Prashant Shenoy}.} \bibinfo{year}{2024}\natexlab{}.
\newblock \showarticletitle{Going Green for Less Green: Optimizing the Cost of Reducing Cloud Carbon Emissions}. In \bibinfo{booktitle}{\emph{Proceedings of the 29th ACM International Conference on Architectural Support for Programming Languages and Operating Systems, Volume 3}} (La Jolla, CA, USA) \emph{(\bibinfo{series}{ASPLOS '24})}. \bibinfo{publisher}{Association for Computing Machinery}, \bibinfo{address}{New York, NY, USA}, \bibinfo{pages}{479–496}.
\newblock
\showISBNx{9798400703867}
\urldef\tempurl%
\url{https://doi.org/10.1145/3620666.3651374}
\showDOI{\tempurl}


\bibitem[Islam et~al\mbox{.}(2018)]%
        {islam2018wace}
\bibfield{author}{\bibinfo{person}{Mohammad~A. Islam}, \bibinfo{person}{Kishwar Ahmed}, \bibinfo{person}{Hong Xu}, \bibinfo{person}{Nguyen~H. Tran}, \bibinfo{person}{Gang Quan}, {and} \bibinfo{person}{Shaolei Ren}.} \bibinfo{year}{2018}\natexlab{}.
\newblock \showarticletitle{Exploiting Spatio-Temporal Diversity for Water Saving in Geo-Distributed Data Centers}.
\newblock \bibinfo{journal}{\emph{IEEE Transactions on Cloud Computing}} \bibinfo{volume}{6}, \bibinfo{number}{3} (\bibinfo{year}{2018}), \bibinfo{pages}{734--746}.
\newblock
\urldef\tempurl%
\url{https://doi.org/10.1109/TCC.2016.2535201}
\showDOI{\tempurl}


\bibitem[Khan et~al\mbox{.}(2018)]%
        {khan2018rapl}
\bibfield{author}{\bibinfo{person}{Kashif~Nizam Khan}, \bibinfo{person}{Mikael Hirki}, \bibinfo{person}{Tapio Niemi}, \bibinfo{person}{Jukka~K Nurminen}, {and} \bibinfo{person}{Zhonghong Ou}.} \bibinfo{year}{2018}\natexlab{}.
\newblock \showarticletitle{Rapl in action: Experiences in using rapl for power measurements}.
\newblock \bibinfo{journal}{\emph{ACM Transactions on Modeling and Performance Evaluation of Computing Systems (TOMPECS)}} \bibinfo{volume}{3}, \bibinfo{number}{2} (\bibinfo{year}{2018}), \bibinfo{pages}{1--26}.
\newblock


\bibitem[Leverich and Kozyrakis(2014)]%
        {leverich2014reconciling}
\bibfield{author}{\bibinfo{person}{Jacob Leverich} {and} \bibinfo{person}{Christos Kozyrakis}.} \bibinfo{year}{2014}\natexlab{}.
\newblock \showarticletitle{Reconciling high server utilization and sub-millisecond quality-of-service}. In \bibinfo{booktitle}{\emph{Proceedings of the Ninth European Conference on Computer Systems}}. \bibinfo{pages}{1--14}.
\newblock


\bibitem[Li et~al\mbox{.}(2023a)]%
        {li2023toward}
\bibfield{author}{\bibinfo{person}{Baolin Li}, \bibinfo{person}{Rohan Basu~Roy}, \bibinfo{person}{Daniel Wang}, \bibinfo{person}{Siddharth Samsi}, \bibinfo{person}{Vijay Gadepally}, {and} \bibinfo{person}{Devesh Tiwari}.} \bibinfo{year}{2023}\natexlab{a}.
\newblock \showarticletitle{Toward Sustainable HPC: Carbon Footprint Estimation and Environmental Implications of HPC Systems}. In \bibinfo{booktitle}{\emph{Proceedings of the International Conference for High Performance Computing, Networking, Storage and Analysis}}. \bibinfo{pages}{1--15}.
\newblock


\bibitem[Li et~al\mbox{.}(2023b)]%
        {li2023making}
\bibfield{author}{\bibinfo{person}{Pengfei Li}, \bibinfo{person}{Jianyi Yang}, \bibinfo{person}{Mohammad~A Islam}, {and} \bibinfo{person}{Shaolei Ren}.} \bibinfo{year}{2023}\natexlab{b}.
\newblock \showarticletitle{Making ai less" thirsty": Uncovering and addressing the secret water footprint of ai models}.
\newblock \bibinfo{journal}{\emph{arXiv preprint arXiv:2304.03271}} (\bibinfo{year}{2023}).
\newblock


\bibitem[Li et~al\mbox{.}(2023c)]%
        {li2023environmentally}
\bibfield{author}{\bibinfo{person}{Pengfei Li}, \bibinfo{person}{Jianyi Yang}, \bibinfo{person}{Adam Wierman}, {and} \bibinfo{person}{Shaolei Ren}.} \bibinfo{year}{2023}\natexlab{c}.
\newblock \bibinfo{title}{Towards Environmentally Equitable AI via Geographical Load Balancing}.
\newblock
\newblock
\showeprint[arxiv]{2307.05494}~[cs.AI]


\bibitem[Liu et~al\mbox{.}(2015)]%
        {liu2015greening}
\bibfield{author}{\bibinfo{person}{Zhenhua Liu}, \bibinfo{person}{Minghong Lin}, \bibinfo{person}{Adam Wierman}, \bibinfo{person}{Steven Low}, {and} \bibinfo{person}{Lachlan L.~H. Andrew}.} \bibinfo{year}{2015}\natexlab{}.
\newblock \showarticletitle{Greening Geographical Load Balancing}.
\newblock \bibinfo{journal}{\emph{IEEE/ACM Transactions on Networking}} \bibinfo{volume}{23}, \bibinfo{number}{2} (\bibinfo{year}{2015}), \bibinfo{pages}{657--671}.
\newblock
\urldef\tempurl%
\url{https://doi.org/10.1109/TNET.2014.2308295}
\showDOI{\tempurl}


\bibitem[Macknick et~al\mbox{.}(2011)]%
        {macknick2011review}
\bibfield{author}{\bibinfo{person}{Jordan Macknick}, \bibinfo{person}{Robin Newmark}, \bibinfo{person}{Garvin Heath}, {and} \bibinfo{person}{KC Hallett}.} \bibinfo{year}{2011}\natexlab{}.
\newblock \showarticletitle{Review of operational water consumption and withdrawal factors for electricity generating technologies}.
\newblock  (\bibinfo{year}{2011}).
\newblock


\bibitem[Macknick et~al\mbox{.}(2012)]%
        {macknick2012operational}
\bibfield{author}{\bibinfo{person}{Jordan Macknick}, \bibinfo{person}{Robin Newmark}, \bibinfo{person}{Garvin Heath}, {and} \bibinfo{person}{Kathleen~C Hallett}.} \bibinfo{year}{2012}\natexlab{}.
\newblock \showarticletitle{Operational water consumption and withdrawal factors for electricity generating technologies: a review of existing literature}.
\newblock \bibinfo{journal}{\emph{Environmental Research Letters}} \bibinfo{volume}{7}, \bibinfo{number}{4} (\bibinfo{year}{2012}), \bibinfo{pages}{045802}.
\newblock


\bibitem[Maji et~al\mbox{.}(2023)]%
        {2023majicarbonmulticloud}
\bibfield{author}{\bibinfo{person}{Diptyaroop Maji}, \bibinfo{person}{Ben Pfaff}, \bibinfo{person}{Vipin P~R}, \bibinfo{person}{Rajagopal Sreenivasan}, \bibinfo{person}{Victor Firoiu}, \bibinfo{person}{Sreeram Iyer}, \bibinfo{person}{Colleen Josephson}, \bibinfo{person}{Zhelong Pan}, {and} \bibinfo{person}{Ramesh~K Sitaraman}.} \bibinfo{year}{2023}\natexlab{}.
\newblock \showarticletitle{Bringing Carbon Awareness to Multi-cloud Application Delivery}. In \bibinfo{booktitle}{\emph{Proceedings of the 2nd Workshop on Sustainable Computer Systems}} (Boston, MA, USA) \emph{(\bibinfo{series}{HotCarbon '23})}. \bibinfo{publisher}{Association for Computing Machinery}, \bibinfo{address}{New York, NY, USA}, Article \bibinfo{articleno}{6}, \bibinfo{numpages}{6}~pages.
\newblock
\showISBNx{9798400702426}
\urldef\tempurl%
\url{https://doi.org/10.1145/3604930.3605711}
\showDOI{\tempurl}


\bibitem[Maps(2024)]%
        {electricitymap}
\bibfield{author}{\bibinfo{person}{Electricity Maps}.} \bibinfo{year}{2024}\natexlab{}.
\newblock \bibinfo{booktitle}{\emph{{Electricity Maps Live 24/7}}}.
\newblock
\urldef\tempurl%
\url{https://app.electricitymaps.com/map}
\showURL{%
\tempurl}


\bibitem[meteologix(2024)]%
        {meteologix}
\bibfield{author}{\bibinfo{person}{meteologix}.} \bibinfo{year}{2024}\natexlab{}.
\newblock \bibinfo{booktitle}{\emph{{meteologix Live 24/7}}}.
\newblock
\urldef\tempurl%
\url{https://meteologix.com/in}
\showURL{%
\tempurl}


\bibitem[Mirhosseini et~al\mbox{.}(2019)]%
        {mirhosseini2019killermicro}
\bibfield{author}{\bibinfo{person}{Amirhossein Mirhosseini}, \bibinfo{person}{Akshitha Sriraman}, {and} \bibinfo{person}{Thomas~F. Wenisch}.} \bibinfo{year}{2019}\natexlab{}.
\newblock \showarticletitle{Enhancing Server Efficiency in the Face of Killer Microseconds}. In \bibinfo{booktitle}{\emph{2019 IEEE International Symposium on High Performance Computer Architecture (HPCA)}}. \bibinfo{pages}{185--198}.
\newblock
\urldef\tempurl%
\url{https://doi.org/10.1109/HPCA.2019.00037}
\showDOI{\tempurl}


\bibitem[Mitchell et~al\mbox{.}(2011)]%
        {mitchell2011pulp}
\bibfield{author}{\bibinfo{person}{Stuart Mitchell}, \bibinfo{person}{Michael OSullivan}, {and} \bibinfo{person}{Iain Dunning}.} \bibinfo{year}{2011}\natexlab{}.
\newblock \showarticletitle{Pulp: a linear programming toolkit for python}.
\newblock \bibinfo{journal}{\emph{The University of Auckland, Auckland, New Zealand}}  \bibinfo{volume}{65} (\bibinfo{year}{2011}).
\newblock


\bibitem[Monserrate(2022)]%
        {monserrate2022cloud}
\bibfield{author}{\bibinfo{person}{Steven~Gonzalez Monserrate}.} \bibinfo{year}{2022}\natexlab{}.
\newblock \showarticletitle{The cloud is material: On the environmental impacts of computation and data storage}.
\newblock  (\bibinfo{year}{2022}).
\newblock


\bibitem[Petrucci et~al\mbox{.}(2015)]%
        {petrucci2015octopus}
\bibfield{author}{\bibinfo{person}{Vinicius Petrucci}, \bibinfo{person}{Michael~A Laurenzano}, \bibinfo{person}{John Doherty}, \bibinfo{person}{Yunqi Zhang}, \bibinfo{person}{Daniel Mosse}, \bibinfo{person}{Jason Mars}, {and} \bibinfo{person}{Lingjia Tang}.} \bibinfo{year}{2015}\natexlab{}.
\newblock \showarticletitle{Octopus-man: Qos-driven task management for heterogeneous multicores in warehouse-scale computers}. In \bibinfo{booktitle}{\emph{2015 IEEE 21st International Symposium on High Performance Computer Architecture (HPCA)}}. IEEE, \bibinfo{pages}{246--258}.
\newblock


\bibitem[Reig(2013)]%
        {wsf}
\bibfield{author}{\bibinfo{person}{Paul Reig}.} \bibinfo{year}{2013}\natexlab{}.
\newblock \bibinfo{title}{What’s the Difference Between Water Use and Water Consumption?}
\newblock \bibinfo{howpublished}{\url{https://www.wri.org/insights/whats-difference-between-water-use-and-water-consumption}}.
\newblock


\bibitem[Reig et~al\mbox{.}(2020)]%
        {reig2020guidance}
\bibfield{author}{\bibinfo{person}{Paul Reig}, \bibinfo{person}{Tianyi Luo}, \bibinfo{person}{Eric Christensen}, {and} \bibinfo{person}{Julie Sinistore}.} \bibinfo{year}{2020}\natexlab{}.
\newblock \showarticletitle{Guidance for calculating water use embedded in purchased electricity}.
\newblock \bibinfo{journal}{\emph{World Resources Institute}} (\bibinfo{year}{2020}).
\newblock


\bibitem[Ritchie et~al\mbox{.}(2024)]%
        {ritchie2024electricity}
\bibfield{author}{\bibinfo{person}{Hannah Ritchie}, \bibinfo{person}{Pablo Rosado}, {and} \bibinfo{person}{Max Roser}.} \bibinfo{year}{2024}\natexlab{}.
\newblock \showarticletitle{Electricity mix}.
\newblock \bibinfo{journal}{\emph{Our World in Data}} (\bibinfo{year}{2024}).
\newblock


\bibitem[Shehabi et~al\mbox{.}(2016)]%
        {shehabi2016united}
\bibfield{author}{\bibinfo{person}{Arman Shehabi}, \bibinfo{person}{Sarah Smith}, \bibinfo{person}{Dale Sartor}, \bibinfo{person}{Richard Brown}, \bibinfo{person}{Magnus Herrlin}, \bibinfo{person}{Jonathan Koomey}, \bibinfo{person}{Eric Masanet}, \bibinfo{person}{Nathaniel Horner}, \bibinfo{person}{In{\^e}s Azevedo}, {and} \bibinfo{person}{William Lintner}.} \bibinfo{year}{2016}\natexlab{}.
\newblock \showarticletitle{United states data center energy usage report}.
\newblock  (\bibinfo{year}{2016}).
\newblock


\bibitem[Siddik et~al\mbox{.}(2021a)]%
        {siddik2021environmental}
\bibfield{author}{\bibinfo{person}{Md~Abu~Bakar Siddik}, \bibinfo{person}{Arman Shehabi}, {and} \bibinfo{person}{Landon Marston}.} \bibinfo{year}{2021}\natexlab{a}.
\newblock \showarticletitle{The environmental footprint of data centers in the United States}.
\newblock \bibinfo{journal}{\emph{Environmental Research Letters}} \bibinfo{volume}{16}, \bibinfo{number}{6} (\bibinfo{year}{2021}), \bibinfo{pages}{064017}.
\newblock


\bibitem[Siddik et~al\mbox{.}(2021b)]%
        {siddik2021datacenterfootprint}
\bibfield{author}{\bibinfo{person}{Md~Abu~Bakar Siddik}, \bibinfo{person}{Arman Shehabi}, {and} \bibinfo{person}{Landon Marston}.} \bibinfo{year}{2021}\natexlab{b}.
\newblock \showarticletitle{The environmental footprint of data centers in the United States}.
\newblock \bibinfo{journal}{\emph{Environmental Research Letters}} \bibinfo{volume}{16}, \bibinfo{number}{6} (\bibinfo{date}{may} \bibinfo{year}{2021}), \bibinfo{pages}{064017}.
\newblock
\urldef\tempurl%
\url{https://doi.org/10.1088/1748-9326/abfba1}
\showDOI{\tempurl}


\bibitem[Souza et~al\mbox{.}(2023)]%
        {souza2023ecovisor}
\bibfield{author}{\bibinfo{person}{Abel Souza}, \bibinfo{person}{Noman Bashir}, \bibinfo{person}{Jorge Murillo}, \bibinfo{person}{Walid Hanafy}, \bibinfo{person}{Qianlin Liang}, \bibinfo{person}{David Irwin}, {and} \bibinfo{person}{Prashant Shenoy}.} \bibinfo{year}{2023}\natexlab{}.
\newblock \showarticletitle{Ecovisor: A Virtual Energy System for Carbon-Efficient Applications}. In \bibinfo{booktitle}{\emph{Proceedings of the 28th ACM International Conference on Architectural Support for Programming Languages and Operating Systems, Volume 2}} (Vancouver, BC, Canada) \emph{(\bibinfo{series}{ASPLOS 2023})}. \bibinfo{publisher}{Association for Computing Machinery}, \bibinfo{address}{New York, NY, USA}, \bibinfo{pages}{252–265}.
\newblock
\showISBNx{9781450399166}
\urldef\tempurl%
\url{https://doi.org/10.1145/3575693.3575709}
\showDOI{\tempurl}


\bibitem[Sukprasert et~al\mbox{.}(2024)]%
        {2024limitationssukprasert}
\bibfield{author}{\bibinfo{person}{Thanathorn Sukprasert}, \bibinfo{person}{Abel Souza}, \bibinfo{person}{Noman Bashir}, \bibinfo{person}{David Irwin}, {and} \bibinfo{person}{Prashant Shenoy}.} \bibinfo{year}{2024}\natexlab{}.
\newblock \showarticletitle{On the Limitations of Carbon-Aware Temporal and Spatial Workload Shifting in the Cloud}. In \bibinfo{booktitle}{\emph{Proceedings of the Nineteenth European Conference on Computer Systems}} (Athens, Greece) \emph{(\bibinfo{series}{EuroSys '24})}. \bibinfo{publisher}{Association for Computing Machinery}, \bibinfo{address}{New York, NY, USA}, \bibinfo{pages}{924–941}.
\newblock
\showISBNx{9798400704376}
\urldef\tempurl%
\url{https://doi.org/10.1145/3627703.3650079}
\showDOI{\tempurl}


\bibitem[Tian et~al\mbox{.}(2019)]%
        {tian2019characterizing}
\bibfield{author}{\bibinfo{person}{Huangshi Tian}, \bibinfo{person}{Yunchuan Zheng}, {and} \bibinfo{person}{Wei Wang}.} \bibinfo{year}{2019}\natexlab{}.
\newblock \showarticletitle{Characterizing and synthesizing task dependencies of data-parallel jobs in alibaba cloud}. In \bibinfo{booktitle}{\emph{Proceedings of the ACM Symposium on Cloud Computing}}. \bibinfo{pages}{139--151}.
\newblock


\bibitem[Treibig et~al\mbox{.}(2010)]%
        {psti}
\bibfield{author}{\bibinfo{person}{J. Treibig}, \bibinfo{person}{G. Hager}, {and} \bibinfo{person}{G. Wellein}.} \bibinfo{year}{2010}\natexlab{}.
\newblock \showarticletitle{LIKWID: A lightweight performance-oriented tool suite for x86 multicore environments}. In \bibinfo{booktitle}{\emph{Proceedings of PSTI2010, the First International Workshop on Parallel Software Tools and Tool Infrastructures}}. \bibinfo{address}{San Diego CA}.
\newblock


\bibitem[Wang et~al\mbox{.}(2024)]%
        {wang2024designing}
\bibfield{author}{\bibinfo{person}{Jaylen Wang}, \bibinfo{person}{Daniel~S. Berger}, \bibinfo{person}{Fiodar Kazhamiaka}, \bibinfo{person}{Celine Irvene}, \bibinfo{person}{Chaojie Zhang}, \bibinfo{person}{Esha Choukse}, \bibinfo{person}{Kali Frost}, \bibinfo{person}{Rodrigo Fonseca}, \bibinfo{person}{Brijesh Warrier}, \bibinfo{person}{Chetan Bansal}, \bibinfo{person}{Jonathan Stern}, \bibinfo{person}{Ricardo Bianchini}, {and} \bibinfo{person}{Akshitha Sriraman}.} \bibinfo{year}{2024}\natexlab{}.
\newblock \showarticletitle{Designing Cloud Servers for Lower Carbon}. In \bibinfo{booktitle}{\emph{ISCA}}.
\newblock
\urldef\tempurl%
\url{https://www.microsoft.com/en-us/research/publication/designing-cloud-servers-for-lower-carbon/}
\showURL{%
\tempurl}


\bibitem[Wang et~al\mbox{.}(2023)]%
        {2023wangpeelingback}
\bibfield{author}{\bibinfo{person}{Jaylen Wang}, \bibinfo{person}{Udit Gupta}, {and} \bibinfo{person}{Akshitha Sriraman}.} \bibinfo{year}{2023}\natexlab{}.
\newblock \showarticletitle{Peeling Back the Carbon Curtain: Carbon Optimization Challenges in Cloud Computing}. In \bibinfo{booktitle}{\emph{Proceedings of the 2nd Workshop on Sustainable Computer Systems}} (Boston, MA, USA) \emph{(\bibinfo{series}{HotCarbon '23})}. \bibinfo{publisher}{Association for Computing Machinery}, \bibinfo{address}{New York, NY, USA}, Article \bibinfo{articleno}{8}, \bibinfo{numpages}{7}~pages.
\newblock
\showISBNx{9798400702426}
\urldef\tempurl%
\url{https://doi.org/10.1145/3604930.3605718}
\showDOI{\tempurl}


\bibitem[Wiesner et~al\mbox{.}(2021)]%
        {wiesner2021waitawhile}
\bibfield{author}{\bibinfo{person}{Philipp Wiesner}, \bibinfo{person}{Ilja Behnke}, \bibinfo{person}{Dominik Scheinert}, \bibinfo{person}{Kordian Gontarska}, {and} \bibinfo{person}{Lauritz Thamsen}.} \bibinfo{year}{2021}\natexlab{}.
\newblock \showarticletitle{Let's wait awhile: how temporal workload shifting can reduce carbon emissions in the cloud}. In \bibinfo{booktitle}{\emph{Proceedings of the 22nd International Middleware Conference}} (Qu\'{e}bec city, Canada) \emph{(\bibinfo{series}{Middleware '21})}. \bibinfo{publisher}{Association for Computing Machinery}, \bibinfo{address}{New York, NY, USA}, \bibinfo{pages}{260–272}.
\newblock
\showISBNx{9781450385343}
\urldef\tempurl%
\url{https://doi.org/10.1145/3464298.3493399}
\showDOI{\tempurl}


\bibitem[Wilkes(2020)]%
        {clusterdata:Wilkes2020}
\bibfield{author}{\bibinfo{person}{John Wilkes}.} \bibinfo{year}{2020}\natexlab{}.
\newblock \bibinfo{title}{Yet more {Google} compute cluster trace data}.
\newblock \bibinfo{howpublished}{Google research blog}.
\newblock
\newblock
\shownote{Posted at \url{https://ai.googleblog.com/2020/04/yet-more-google-compute-cluster-trace.html}.}.


\bibitem[Wu et~al\mbox{.}(2022)]%
        {wu2022sustainable}
\bibfield{author}{\bibinfo{person}{Carole-Jean Wu}, \bibinfo{person}{Ramya Raghavendra}, \bibinfo{person}{Udit Gupta}, \bibinfo{person}{Bilge Acun}, \bibinfo{person}{Newsha Ardalani}, \bibinfo{person}{Kiwan Maeng}, \bibinfo{person}{Gloria Chang}, \bibinfo{person}{Fiona Aga}, \bibinfo{person}{Jinshi Huang}, \bibinfo{person}{Charles Bai}, \bibinfo{person}{Michael Gschwind}, \bibinfo{person}{Anurag Gupta}, \bibinfo{person}{Myle Ott}, \bibinfo{person}{Anastasia Melnikov}, \bibinfo{person}{Salvatore Candido}, \bibinfo{person}{David Brooks}, \bibinfo{person}{Geeta Chauhan}, \bibinfo{person}{Benjamin Lee}, \bibinfo{person}{Hsien-Hsin~S. Lee}, \bibinfo{person}{Bugra Akyildiz}, \bibinfo{person}{Maximilian Balandat}, \bibinfo{person}{Joe Spisak}, \bibinfo{person}{Ravi Jain}, \bibinfo{person}{Mike Rabbat}, {and} \bibinfo{person}{Kim Hazelwood}.} \bibinfo{year}{2022}\natexlab{}.
\newblock \showarticletitle{Sustainable ai: Environmental implications, challenges and opportunities}.
\newblock \bibinfo{journal}{\emph{Proceedings of Machine Learning and Systems}}  \bibinfo{volume}{4} (\bibinfo{year}{2022}), \bibinfo{pages}{795--813}.
\newblock


\bibitem[Zhan et~al\mbox{.}(2017)]%
        {zhan2017parsec3}
\bibfield{author}{\bibinfo{person}{Xusheng Zhan}, \bibinfo{person}{Yungang Bao}, \bibinfo{person}{Christian Bienia}, {and} \bibinfo{person}{Kai Li}.} \bibinfo{year}{2017}\natexlab{}.
\newblock \showarticletitle{PARSEC3. 0: A multicore benchmark suite with network stacks and SPLASH-2X}.
\newblock \bibinfo{journal}{\emph{ACM SIGARCH Computer Architecture News}} \bibinfo{volume}{44}, \bibinfo{number}{5} (\bibinfo{year}{2017}), \bibinfo{pages}{1--16}.
\newblock


\bibitem[Zhang et~al\mbox{.}(2022)]%
        {zhang2022schedinspector}
\bibfield{author}{\bibinfo{person}{Di Zhang}, \bibinfo{person}{Dong Dai}, {and} \bibinfo{person}{Bing Xie}.} \bibinfo{year}{2022}\natexlab{}.
\newblock \showarticletitle{Schedinspector: A batch job scheduling inspector using reinforcement learning}. In \bibinfo{booktitle}{\emph{Proceedings of the 31st International Symposium on High-Performance Parallel and Distributed Computing}}. \bibinfo{pages}{97--109}.
\newblock


\bibitem[Zuccon et~al\mbox{.}(2023)]%
        {zuccon2023beyond}
\bibfield{author}{\bibinfo{person}{Guido Zuccon}, \bibinfo{person}{Harrisen Scells}, {and} \bibinfo{person}{Shengyao Zhuang}.} \bibinfo{year}{2023}\natexlab{}.
\newblock \showarticletitle{Beyond CO2 emissions: The overlooked impact of water consumption of information retrieval models}. In \bibinfo{booktitle}{\emph{Proceedings of the 2023 ACM SIGIR International Conference on Theory of Information Retrieval}}. \bibinfo{pages}{283--289}.
\newblock


\end{thebibliography}
